\documentclass[fleqn,usenatbib]{mnras}

\usepackage{newtxtext,newtxmath}

\usepackage[T1]{fontenc}

\DeclareRobustCommand{\VAN}[3]{#2}
\let\VANthebibliography\thebibliography
\def\thebibliography{\DeclareRobustCommand{\VAN}[3]{##3}\VANthebibliography}

\usepackage{graphicx}	
\usepackage{bm}
\usepackage{makecell}
\usepackage{placeins}
\usepackage{float}

\title[Merger-induced galaxy transformations]{Merger-induced galaxy transformations in the \texttt{ARTEMIS} simulations}
\author[A. M. Dillamore et al.]{
Adam M. Dillamore$^{1}$\thanks{E-mail: amd206@cam.ac.uk},
Vasily Belokurov,$^{1,2}$
Andreea S. Font$^{3}$
and Ian G. McCarthy$^{3}$
\\
$^{1}$Institute of Astronomy, University of Cambridge, Madingley Road, Cambridge CB3 0HA, UK\\
$^{2}$Institut de Ci\`encies del Cosmos (ICCUB), Universitat de Barcelona (IEEC-UB), Mart\'i Franqu\`es 1, E08028 Barcelona, Spain\\
$^{3}$Astrophysics Research Institute, Liverpool John Moores University, 146 Brownlow Hill, Liverpool L53RF, UK
}

\date{Accepted XXX. Received YYY; in original form ZZZ}

\pubyear{2021}

\begin{document}
\label{firstpage}
\pagerange{\pageref{firstpage}--\pageref{lastpage}}
\maketitle

\begin{abstract}
Using the \texttt{ARTEMIS} set of 45 high-resolution cosmological simulations, we investigate a range of merger-induced dynamical transformations of Milky Way-like galaxies. We first identify  populations of accreted stars on highly radial orbits, similar to the `\textit{Gaia} Sausage' in the Milky Way. We show that $\approx1/3$ of the \texttt{ARTEMIS} galaxies contain a similar feature, and confirm that they usually comprise stellar debris from the most massive accreted satellite. Selecting 15 galaxies with discs at the present-day, we study their changes around the times of the GS-like mergers. Dark matter haloes of many of these exhibit global changes in shape and orientation, with almost half becoming significantly more spherical when the mergers occur. Focusing on the galaxies themselves, we find that 4/15 have stellar discs which experience large changes in the orientation of their angular momentum (AM) axes, at rates of up to $\sim60$ degrees Gyr$^{-1}$. By calculating the orbital angular momentum axes of the satellites before they are accreted, we show that there is a tendency for the disc's AM to become more aligned with this axis after the merger. We also investigate the origin of \textit{in situ} retrograde stars, analogous to the `Splash' in the Milky Way. Tracing them back to earlier snapshots, we demonstrate that they were often disrupted onto their extreme orbits by multiple early mergers. We also find that the total mass of these stars outside the central regions positively correlates with the total accreted stellar mass.
\end{abstract}

\begin{keywords}
Galaxy: formation -- Galaxy: kinematics and dynamics -- Galaxy: structure -- Galaxy: halo
\end{keywords}



\section{Introduction}
\label{section:introduction}

As with the rest of the Universe, the Milky Way’s most transformative years have long passed \citep[][]{Madau2014}. Only with the help of a stellar archaeological record can we hope to reconstruct the sequence of events that unfolded many Gyrs ago and gave the Galaxy its shape and form. The stellar halo contributes an insignificant mass to the total Galactic budget \citep[e.g.,][]{Bell2008,Deason_mass,Mackereth} but serves as a reliable repository of historical information. Cosmological hydrodynamical simulations suggest two main stellar halo formation channels: accreted and in-situ \citep[see e.g.,][]{Cooper2010,Cooper2015,Font2011,tissera2014,pillepich2015}. The accreted stellar halo component is built up through disruption and mixing of smaller sub-galactic fragments \citep[see][]{KVJ1996,KVJ1998,Helmi1999,BJ2005}, while the main pathway to the in-situ halo creation is via the heating of the pre-existing disc \citep[e.g.,][]{Zolotov2009,McCarthy}. Note that only a very small number of relatively massive ($10^8<M_*<10^9 \, {\rm M}_{\odot}$) dwarf satellites are predicted to contribute the bulk of the accreted stellar halo component \citep[e.g.,][]{Font2006,Delucia2008,Deason_break,D'Souza2018,elias2018,monachesi2019}. Over the last three decades, such an ancient massive merger event has been repeatedly invoked as the principal solution to the puzzle of the Galactic thick disc and the observed $\alpha$-[Fe/H] bi-modality \citep[see e.g.,][]{Gilmore1989,Chiappini1997,Villalobos2008}. Yet no clear sign of this long-suspected catastrophic incident had been detected, that is until recently. 

Recent observations from the \textit{Gaia} space telescope \citep{Gaia} have significantly improved the accuracy and quantity of Milky Way astrometric data. These have been combined with various spectroscopic observations, with the result that stellar positions and velocities are now known with far greater precision than a decade ago. For example, using data from \textit{Gaia} DR1, \citet{Belokurov} showed that the local stellar halo of the Milky Way (MW) contains a large contribution from stars on highly radial orbits with little angular momentum, dubbed the `\textit{Gaia} Sausage' (GS). The authors concluded that these stars originated in a large satellite galaxy (of total mass $>10^{10}\, {\rm M}_\odot$) which underwent a major merger with the MW \mbox{8-10 Gyr ago.} The nature of the ancient merger was soon confirmed with the Gaia DR2 data \citep{Helmi,Haywood2018,Deason2018,Mackereth2019}. The follow-up studies revealed that the GS merger not only dictates the structure of the inner stellar halo \citep[see e.g.,][]{Myeong,Myeong2018action,Myeong2018gc,Koppelman2018,Lancaster2019,Iorio,Simion2019,Bird2019}, but appears to be contemporaneous with the demise of the thick disc, the emergence of the in-situ halo and the formation of the bar. \citep[][]{Dimatteo2019,Fantin2019,Gallart2019,splash,mash,Bonaca2020,Fragkoudi,Sit2020}.

The ancient GS merger played a pivotal role in the transformation of our Galaxy. Here we study simulated analogues of this event in the \texttt{ARTEMIS} suite to elucidate the response it triggered in the young Milky Way. While several studies have already looked into the properties of the progenitor dwarf galaxy, so far little attention has been given to the multitude of the effects of the interaction on the host. \citet{Fattahi} used 28 galaxies in the Auriga suite \citep[see][]{Auriga} to search for GS-like populations. They found that roughly a third of the galaxies contained such a feature which dominated the stellar halo, and confirmed that these stars were delivered in one or two mergers with large dwarf galaxies of stellar mass $\gtrsim 10^{8}\, {\rm M}_\odot$. The prevalence and the properties of the simulated GS encounter found by \citet{Fattahi} is in general agreement with other studies based on cosmological galaxy formation simulations \citep[e.g.,][]{Belokurov,Mackereth2019,Bignone2019,Elias2020}. Note, however, that several of the above studies use implicit or explicit constraints on the accretion history of the MW look-alike. In this work we dispense with such a bias, analyzing galaxies that obey solely the MW mass constraint.

The flooding of the Milky Way with the GS tidal debris is only a part of the metamorphosis the stellar halo underwent. Gaia DR1 and DR2 data were used to uncover a substantial population of stars with metallicities typical of the Galactic disc but on highly eccentric or retrograde orbits \citep[see][]{Bonaca2017,Gallart2019,Dimatteo2019,splash}. The exact origin of this population remains uncertain, but all of the above studies agree that these stars were born inside the MW, before or around the time of the GS accretion. These stars could be, for example, the remnants of the Galactic ancient disc splashed by the violent interaction between the MW and the GS progenitor. To this end, \citet{mash} used simulated galaxies in Auriga to study analogues of the GS merger and its connection to the disc and stellar halo of the host. They showed that stars belonging to retrograde `Splash' populations were indeed formed in the proto-discs and scattered onto extreme orbits by gas-rich mergers. Fresh gas supply provided by the infalling massive dwarf and the intense nature of the ensuing interaction additionally lead to a star-burst in the central MW contributing both to the thick disc and the Splash components \citep[see][]{mash}. Under the assumption that the Splash is created out of the Galactic proto-disc heated up by the GS merger, its stellar mass and kinematics can be used to gauge the mass ratio of the in-falling satellite and the host \citep[][]{splash,mash}. 

The GS merger punctuates one of the busiest epochs in the MW's life, the period expected to witness the transformation of the DM halo itself. The details of the galactic dark matter distribution, including the shape of the DM halo and the properties of the small-scale substructure such as dwarf satellites, are controlled by the host halo’s accretion history which is in turn connected to the characteristics of the surrounding large scale structure \citep[see e.g.,][]{Frenk1988, Cole1996, Libeskind2005, Allgood2006, Hahn2007, VeraCiro2011, Codis2015}. While the Milky Way is universally considered to be the best-equipped dark matter “laboratory”, the observational progress to measure the Galactic DM distribution has been painfully slow. In the Solar neighborhood, strong constraints exist as to the DM halo shape and the local density normalisation \citep[e.g.,][]{Piffl2014, Nitschai2020}, but further from the Sun our vision of the DM halo remains blurred. For example, only a few kpc away, the DM halo has been claimed to have a spherical \citep[e.g.,][]{Wegg2019}, an oblate \citep[e.g.,][]{Olling2000} and a prolate \citep[e.g.,][]{Bowden2016} shape. The most remarkable claim has been perplexing and galvanising the community for the last decade: according to \citet{Law}, who used the Sagittarius stellar stream to measure the large-scale DM halo shape, the MW’s halo is triaxial with its minor axis in the Galactic disc and the major one aligned with the orbital plane of the Magellanic Clouds. This measurement is extraordinary because, while it is true that stellar streams are the most robust accelerometers available locally \citep[see e.g.,][]{Koposov2010, Bowden2015, Bonaca2018}, the DM halo configuration inferred is not dynamically stable \citep[][]{Debattista2013}. Note that, as has been recently realised, all of the previous Galactic DM shape inference ought to be re-evaluated now to take into account the interaction between the MW and the Large Magellanic Cloud (LMC) \citep[e.g.,][]{Garavito2019,Erkal2019,Vasiliev2021}.

One well-established observation which hints at a very particular arrangement of the MW’s dark matter is the tendency of its dwarf satellites to orbit in a relatively tight polar (i.e., set up perpendicular to the Galactic disc) disc-like pattern \citep[e.g.,][]{LyndenBell1976, Kroupa2005}. Comparison with cosmological galaxy formation simulations suggests that the dwarf satellite motions tend to align with the outer portions of the host’s DM halo while the Galaxy itself reshapes and re-aligns the inner halo’s DM distribution \citep[][]{Deason_mismatch,Tenneti2014,Shao2016}. Note, however, that satellites also prefer to be co-planar with the host’s stellar disc, making the MW’s arrangement rather remarkable and rare. A picture therefore emerges in which the main direction of the Galactic DM accretion underwent a dramatic swerve in the last 8-10 Gyr instigated by a change in the surrounding large-scale structure \citep[see][]{Shao2021}. The GS accretion is probably the only major event in our Galaxy’s past that occurred at approximately the same time. Even more strikingly, it has been recently demonstrated that the tidal stellar debris cloud it left behind is closely aligned with the disk of satellites and the orbit of the Magellanic Clouds \citep[][]{Iorio}. \citet{Fattahi} demonstrated that while dominating the stellar halo, the GS progenitor was not an important contributor to the MW’s dark matter halo \citep[see also][]{Necib2019,Evans2019,Bozorgnia2020}. However, as the largest dwarf galaxies are accreted along the principal feeding filaments, the GS merger instead was likely a harbinger of the ensuing switch in the Galactic DM accretion.

Connected to the global drastic re-arrangement of the baryonic and dark matter distribution is the possibility of a significant change in the direction of the net angular momentum (AM) vector of the Galaxy's baryons, i.e., its stellar and gaseous discs \citep[][]{Debattista2008,Debattista2015}. The AM evolution can be fast and furious (spin flipping) and/or slow and slight (spin precession). Previously, work on such a merger-induced spin re-alignment has often been presented in the context of large-scale filamentary structure. \citet{Welker} showed using cosmological simulations that massive galaxies which have undergone major mergers tend to have a spin axis perpendicular to their nearest filament. They found that this realignment is driven by the mergers, with higher mass ratio mergers resulting in a greater `memory loss of the post-merger spin direction'. However, low-mass galaxies that experience no significant mergers generally have an AM axis aligned with the filament. \citet{Kraljic} have similarly found in simulations that galaxies with a stellar mass below $\sim10^{10}\, {\rm M}_\odot$ have spins aligned with nearby filaments, while higher mass galaxies tend to be orthogonal. Observational evidence for this mass dependence was found by \citet{Welker_2}. Using data from the SAMI Galaxy Survey they show that the stellar transition mass between these two behaviours is between $10^{10.4}$ and $10^{10.9}\, {\rm M}_\odot$. This is comparable to recent estimates of MW's stellar mass of $\sim10^{10.8}\, {\rm M}_\odot$ \citep[e.g.,][]{Licquia, McMillan}. It is therefore plausible that our galaxy underwent such a spin flip caused by the merger with the GS progenitor. Determining whether this occurred is of great importance for reconstructing the GS merger. Recent studies such as \citet{Naidu_reconstruction} have used a combination of surveys and N-body simulations to deduce the likely geometry of this ancient merger \citep[see also][]{Vasiliev_radial}. In order to fully understand this, it is necessary to consider whether the MW itself changed orientation during the merger.

This work uses the \texttt{ARTEMIS} set of 45 zoomed-in, hydrodynamical galaxy formation simulations \citep[see][]{Font}. These are high resolution simulations, with a baryon particle mass of order~$10^4\, {\rm M}_\odot$, similar to that used by the Auriga Project. However, \texttt{ARTEMIS} has some advantages over Auriga for the purposes of our study: the suite is 50\% larger (Auriga has 30 MW-type galaxies); and the selection of host haloes is based only on total mass, whereas in Auriga an additional constraint was imposed that the haloes must be relatively isolated \citep[see][]{Auriga}. By including systems that experienced recent major mergers, we can investigate the multitude of pathways in which MW-mass galaxies can form, and estimate the prevalence of GS-like progenitors without any assumptions about the host merger histories.\par

The aim of the work is to address a variety of dynamical transformations which MW analogues can undergo as a result of massive mergers. We largely focus on the changes in stellar kinematics and spatial distributions throughout the history of the galaxies, particularly around the times of mergers. We aim to address whether spin flips do occur in the MW analogues, and whether they coincide with these major mergers. Of particular importance is whether the stellar discs are destroyed as a result, since this has implications for the evolution of the galaxy's morphology. We also investigate whether the directions of the flips are connected to the merger geometry. Following on from the earlier work \citep{splash,mash,Dimatteo2019} we probe the origin of the retrograde `Splash' populations. In particular, we examine whether these stars belonged to a disc prior to any merger which may have disrupted them. 

Our survey of the variety of host transformations arising as a result of its interaction with a massive in-falling satellite is inspired by the ancient GS event in our Galaxy as uncovered recently with the use of the {\it Gaia} data. Such major mergers often lead to a rapid satellite radialization \citep[see][]{Vasiliev_radial} spraying, as a result, the satellite's debris onto highly eccentric orbits in the host's inner halo. Motivated by this, we use the presence of a significant local population with acute radial anisotropy as the principal marker of such a past major merger. We choose not to add additional observational constraints such as the exact epoch of the accretion or the mass of the satellite even though some new pieces of evidence have been unearthed recently. By not using the GS event as a rigid template for our study we gain access to a wider variety of assembly histories and a richer spectrum of host transformations. Note, however, that throughout the analysis we do take care to point out a handful of the closest analogs of the MW-GS interaction. While the ARTEMIS suite does present us with an opportunity to study MW-mass hosts without a morphological type constraint, in the present work we focus predominantly on the hosts that appear disc-like at redshift $z=0$.

The structure of the paper is as follows. Section \ref{sec:sims} summarises the \texttt{ARTEMIS} simulations and the main properties of the Milky Way-mass hosts. Section \ref{sec:GS} investigates the velocity distributions of accreted stars in the Solar neighborhood (Section \ref{section:velocity_space}) and the process of identifying GS-like progenitors of these galaxies in these regions using  kinematic signatures (Section \ref{section:selection}), discussing also the formation histories of galaxies with GS-like features (Section \ref{section:properties}). Section \ref{sec:host_transformations} focuses primarily on two main types of galaxy transformations that accompany these massive mergers, namely disc flipping and disc splashing. The principal findings and conclusions are summarised in Section \ref{sec:concl}. A table containing the most important parameters for the galaxies and mergers studied is also presented in \ref{sec:table}.

\section{Simulations}
\label{sec:sims}

\texttt{ARTEMIS} is a suite of $45$ zoomed hydrodynamical simulations of Milky Way-mass galaxies set in a WMAP-9 $\Lambda$CDM cosmology. The simulations were carried out with the Gadget-3 code with an updated hydro solver, last described in \citet{springel2005}. The subgrid physical prescriptions include model metal-dependent radiative cooling in the presence of a photo-ionizing UV background, star formation, supernova and active galactic nuclei feedback, stellar and chemical evolution and formation of black holes. More details about their implementation in the EAGLE simulations are given in \citet{schaye2015}, \citet{crain2015} and references therein. The prescriptions are unchanged in ARTEMIS, with the exception of the stellar feedback which was re-calibrated in order to re-gain the good match to the observed stellar mass -- halo mass relation of galaxies obtained with EAGLE (Note that the stellar feedback is calibrated, by construction, against this set of observables; Also, as shown in \citet{schaye2015},  whenever the mass resolution is increased, the stellar feedback needs to be adjusted to recover the good match to the calibration observables. This is also needed for ARTEMIS, which achieves a mass resolution higher by a factor of $\sim 7$ than the highest resolution EAGLE runs; see \citealt{Font} for details.).

The 45 systems were initially selected from a 25 Mpc/$h$ periodic box collisionless simulation based solely on their total mass (i.e., no additional constraints on merger histories). The total masses range between $8 \times 10^{11} < {\rm M}_{200}/{\rm M}_{\odot} < 2 \times 10^{12}$, where ${\rm M}_{200}$ is the mass enclosing a mean density of $200$ times the critical density at present time. The particle masses are $1.17\times10^5$ M$_\odot\/h^{-1}$ for dark matter and $2.23\times10^4$ M$_\odot\/h^{-1}$ for the initial baryon mass, while the Plummer-equivalent softening is $125$~pc $h^{-1}$. Gravitationally bound haloes and subhaloes are identified using the \texttt{SUBFIND} algorithm of \citet{dolag2009}.

The simulations match the stellar mass -- halo mass relation of galaxies by construction, but they match a series of global and structural properties of MW-mass galaxies without further tuning, e.g., the sizes, stellar masses, luminosities and average metallicities ([Fe/H]), as well as the spatial distribution of metals \citep{Font}. Furthermore, a broad series of properties of satellite dwarf galaxies of MW-mass systems are also well matched, for example, the luminosity functions, radial distributions and quenched fractions of dwarf galaxies in the MW, M31 and other MW analogues \citep{font2021,font_q}. More details about the properties of MW-mass hosts and of their present-day satellites can be found in the respective studies.

We also constructed the merger trees for all MW-mass hosts, using the methods described in \citet{mcalpine2016}. We use these to identify the most massive acrreted progenitor of each host MW-type galaxy, their times of accretion and to follow these progenitors (and their hosts) backwards and forwards in time.

\section{Prevalence of GS-like features}
We begin by addressing whether the \texttt{ARTEMIS} galaxies contain GS-like features, and if so, how frequently occurring they are. In Section~\ref{section:velocity_space} we find such features using the velocity space of accreted stars. We determine their incidence in Section~\ref{section:selection} and select a sample of galaxies containing GS analogues for further study. The evolution of these galaxies is investigated in Section~\ref{section:properties}, including their accretion histories.
\label{sec:GS}
\subsection{Velocity space distributions of accreted stars}\label{section:velocity_space}
\begin{figure}
    \centering
    \includegraphics[width=\columnwidth]{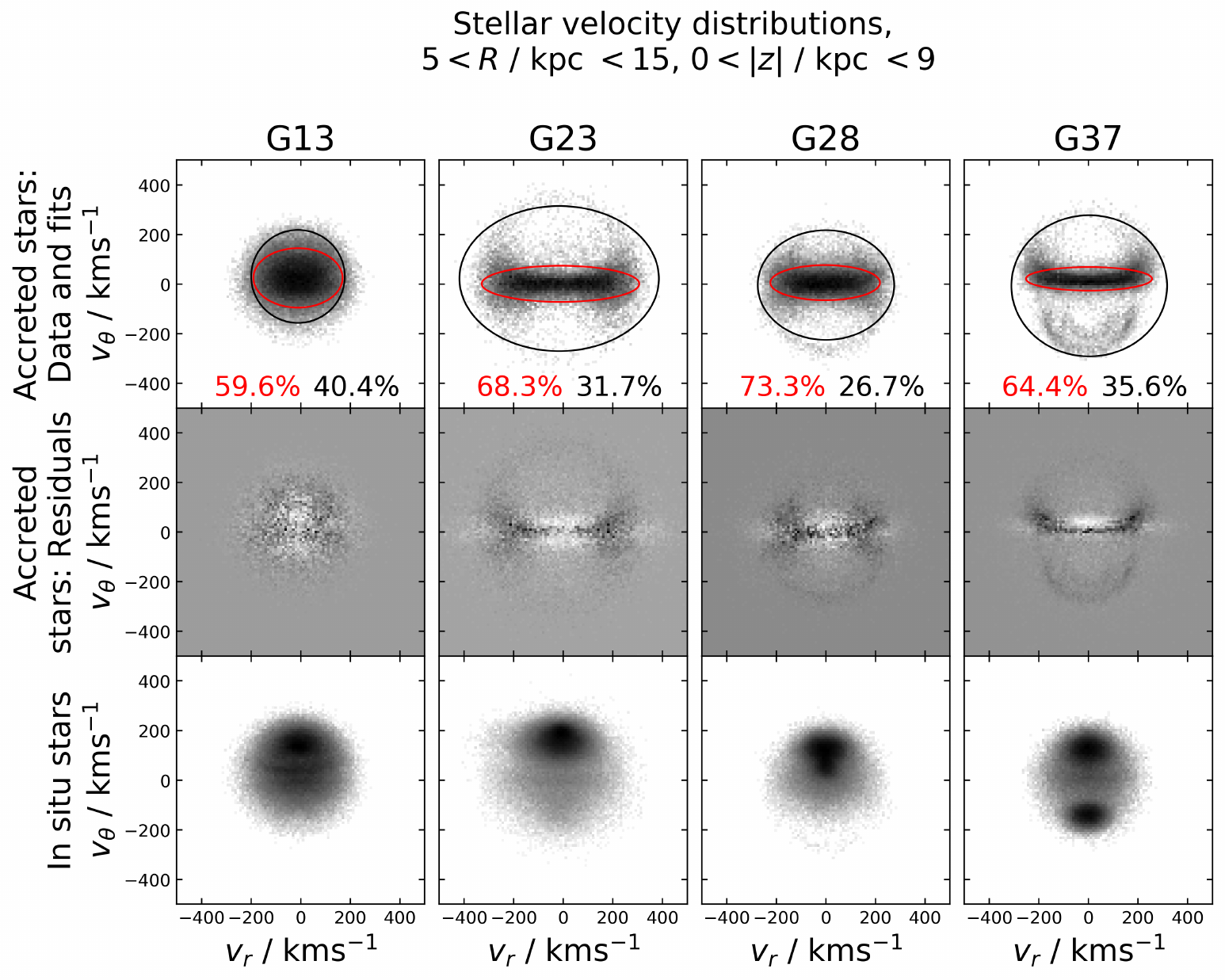}
    \caption{Velocity distributions of accreted star particles (top row), residuals of the Gaussian fits (middle row), and velocity distributions of \textit{in situ} stars (bottom row). The semi-axis of each ellipse along the $i$th direction is equal to 2$\sigma_i$, where $\sigma_i^2$ is the variance of the fitted Gaussian along that direction. The red ellipse corresponds to the more radially anisotropic component, and the black ellipse to the more isotropic component. The corresponding weights are shown as percentages. Black pixels in the residual plots correspond to an excess of stars compared to the fits, and white pixels to a depletion. G23, G28 and G37 each contain a highly radially anisotropic feature, while G13 is more isotropic in comparison. The residuals show excesses of star particles at high $|v_r|$ and low $|v_\theta|$ compared to the fits. In G23 and G28 the high concentrations of \textit{in situ} stars at $v_\theta\sim200$~kpc correspond to stellar discs (see Section~\ref{section:selection}). G37 also contains a similar feature at negative $v_\theta$.}
    \label{fig:fits}
\end{figure}
To identify approximately how many \texttt{ARTEMIS} galaxies contain a feature resembling the GS, we followed a similar method to \citet{Fattahi}. We used a coordinate system centred on the host's centre of potential (CoP), with the $z$-axis aligned with the galaxy's angular momentum vector. Following \citet{Fattahi}, this was chosen to be the total angular momentum of all star particles within a radius of $r=10$~kpc. We selected stars at redshift $z=0$ bound to the most massive (host) halo which satisfied the cuts:
\begin{itemize}
    \item cylindrical radii $R$ in the range $5<R$~/~kpc~$<15$;
    \item heights above the galactic plane in the range $0<|z|$~/~kpc~$<9$;
    \item flagged as accreted only.
\end{itemize}
The spatial cuts roughly mimic the volume in which \textit{Gaia} observes stars in the Solar neighbourhood \citep[see Fig. 1 in][]{Belokurov}. Similarly to \citet{Fattahi}, the purpose of these cuts is not to accurately simulate the observations of \textit{Gaia}, so a more realistic cut is not required. Star particles flagged as \textit{in situ} are those born bound to the host subhalo. This is defined as the most massive subhalo in the friends-of-friends group. Star particles flagged as accreted are all those not bound to this host subhalo at their birth \citep[see][]{Font}. Choosing only accreted particles allows the minimum value of $|z|$ to be reduced to 0 kpc without the sample being contaminated by an \textit{in situ} disc.\par
To quantify the presence of GS-like features, we computed the velocity components of star particles in spherical polar coordinates ($v_r$, $v_\phi$, $v_\theta$). Here $r$ is the spherical galactocentric radius, $\phi$ is the polar angle measured from the positive $z$-axis, and $\theta$ is the azimuthal angle measured from an arbitrary reference plane. Following \citet{Belokurov} we plotted 2D histograms of $v_r$ against $v_\theta$ for the star particles in each sample. Using the expectation-maximization algorithm \texttt{GaussianMixture} from the \texttt{scikit-learn} library \citep{scikit-learn}, we fitted Gaussians to the 3D velocity distributions in spherical polars. A two-component fit was used, with initializations chosen such that the components represented a) any radially anisotropic population present, and b) a more isotropic population. In the Milky Way, the radially anisotropic component is the GS. Similarly to \citet{Fattahi}, we characterised this component by two quantities:
\begin{itemize}
    \item proportional contribution to the accreted population;
    \item anisotropy parameter, defined as $\beta=1-\frac{\sigma_\theta^2+\sigma_\phi^2}{2\sigma_r^2}$. Here $\sigma_i$ is the velocity dispersion along the $i$th direction. For a population of stars on purely radial orbits, $\beta=1$. If the velocity distribution is isotropic, $\beta=0$.
\end{itemize}
We tested the fitting process by applying it to mock data drawn from a combination of two 3D Gaussian distributions, with a variety of covariance matrices and relative weights. In nearly all cases the fitting converged on the correct covariances with reasonably good accuracy. However, the fitted weight values were sometimes inaccurate, with errors of up to $\sim0.2$. This occurred in cases where the contribution and $\beta$ of the more radially anisotropic component were both small. Hence the estimates for contribution may be somewhat inaccurate in some cases, and should only be considered as approximate guides.\par

The quality of the fits was checked by plotting and visually inspecting the corresponding ellipses in $v_\theta/v_r$ and $v_\phi/v_r$ projections of velocity space, along with the residuals. Examples from four different galaxies can be seen in Fig.~\ref{fig:fits}. These were selected to show a variety of appearances. Galaxies G23, G28 and G37 each have conspicuous radially anisotropic components which dominate the distributions, while the G13 distribution is more isotropic.\par
In a large majority of cases any visible radially anisotropic features are fitted moderately well by the corresponding Gaussian components (red ellipses), such as in G23 and G28. However, the plots of residuals reveal two major discrepancies between the data and the fits:
\begin{itemize}
    \item In cases where a prominent radially anisotropic component is present, the fit over-predicts at small $|v_r|$ and under-predicts at higher $|v_r|$. This same non-Gaussian nature of the distribution was observed in the GS by \citet{Belokurov}, and is due to an excess of stars with large apo-centres \citep[][]{Deason2018} that can be linked to the orbital radialization of the GS progenitor \citep[][]{Vasiliev_radial}. As a consequence, the phase-space distribution of the tidal debris is strongly bi-modal in the radial velocity dimension \citep[see also][]{Lancaster2019,Necib2019,Iorio, Iorio2021}.
    \item More complex deviations from Gaussian distributions are seen, for instance in G37. For example, there are loops of higher stellar density at large $|v_\theta|$, which are difficult to fit accurately with any simple model. In some cases these may further reduce the reliability of the contribution estimates.
\end{itemize}
The first point suggests that a more suitable model for some of these velocity distributions would involve additional Gaussians. These would be offset from zero, representing the excesses of stars at high $|v_r|$. However, for this work only approximate values for the parameters are required to gain an estimate of the prevalence of GS-like features. Hence this simple model is deemed adequate for these purposes.

\subsection{Selection of galaxies containing GS-like features}
\label{section:selection}
\begin{figure}
    \centering
    \includegraphics[width=0.9\columnwidth]{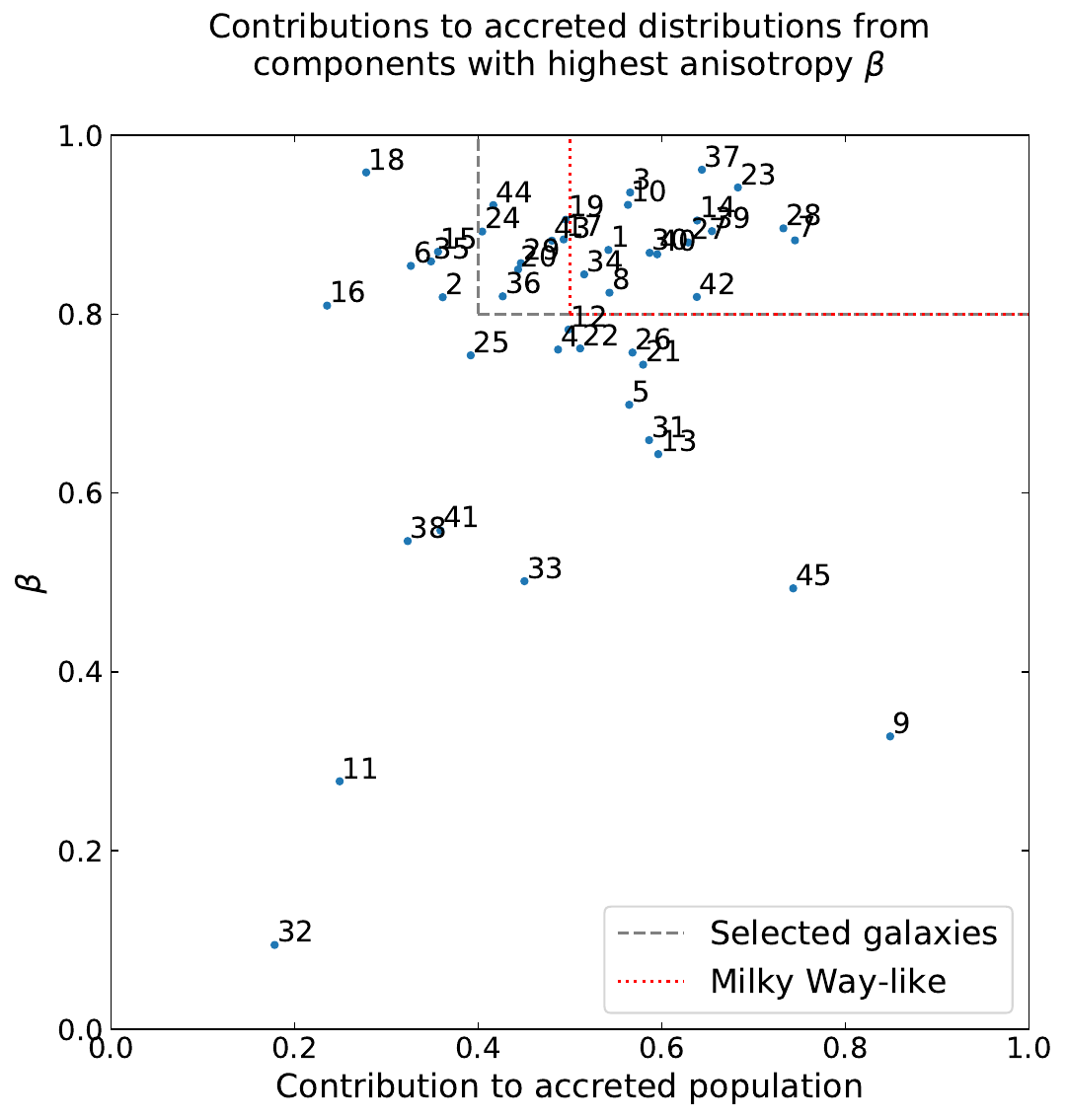}
    \caption{Anisotropy parameter $\beta$ and contribution to the accreted population of the more radially anisotropic fitted components. The region enclosed by the red dotted lines is considered Milky Way-like, whereas the grey dashed lines mark the region from which an initial 23 galaxies were selected (Sample~GS/all). A large proportion of \texttt{ARTEMIS} galaxies have a strongly radial component, with about $2/3$ having $\beta>0.8$. Only in some of these does this component dominate; in total $\sim1/3$ of the galaxies have both $\beta>0.8$ and a contribution of $>50\%$, similar to the Milky Way.}
    \label{fig:beta_cont}
\end{figure}

The procedure outlined above was used to find estimates of $\beta$ and contribution for the components with highest $\beta$ in all 45 galaxies. These are plotted in Fig.~\ref{fig:beta_cont}.

The corresponding values for the GS in the Milky Way are $\beta\gtrsim 0.8$ and a contribution of greater than 50\% \citep{Belokurov, Fattahi}. Fig.~\ref{fig:beta_cont} shows that $\sim30$ of the 45 \texttt{ARTEMIS} galaxies have features with estimated values of $\beta>0.8$. Fifteen of these also have an estimated contribution \hbox{of $>50\%$}, which is $1/3$ of the \texttt{ARTEMIS} galaxies. This is in agreement with the proportion of Auriga galaxies with a similar feature found by \citet{Fattahi}, albeit using different selection cuts on the star particles.\par
From this distribution we selected a sample of galaxies containing a feature comparable to the GS. Let this be denoted Sample~GS/all. We chose the requirement for Sample~GS/all to be $\beta>0.8$ and contribution $>40\%$. The constraint on contribution was relaxed from $>50\%$ due to the uncertainties in these values, and to increase the sample size. Twenty three galaxies survived this cut. Again, we inspected the velocity distributions to visually confirm that each of the selected galaxies contained a strongly anisotropic feature which dominated the distribution.\par
Since some \texttt{ARTEMIS} galaxies are not discs, we imposed a further cut to remove these from the selection. This was based on the co-rotation parameter \citep[see][]{Correa}, which we define as

\begin{equation}\label{equation:kappa}
    \kappa_\mathrm{co} = \frac{1}{K_{30\,\mathrm{kpc}}}\sum_{\substack{i \\ v_\theta>0}}^{r<30\,\mathrm{kpc}}\frac{1}{2}m_iv_{\theta,i}^2.
\end{equation}

This is the fraction of stellar kinetic energy associated with ordered motion about the same axis \emph{and in the same sense} as the galaxy's rotation. Here $m_i$ and $v_{\theta,i}$ are the mass and azimuthal velocity of the $i$th star particle, and the sum is over all star particles with $v_\theta>0$ and galactocentric radius $r<30$ kpc. $K_{30\,\mathrm{kpc}}$ is the total kinetic energy of star particles within $r=30$ kpc. This is a physical scale often used in simulations (see, e.g., \citealt{schaye2015}), as it closely mimics Petrosian magnitude-based estimates implemented in standard pipelines for estimating the galaxy stellar mass function (e.g., SDSS). For MW-mass galaxies, it also encompasses most of the stellar mass of the system (see also Fig.~1 in \citealt{Font} for the match to the observed galaxy size -- stellar mass relation, where stellar masses were computed within the $30$~kpc aperture). Note that all distances quoted throughout this paper are physical, not comoving.\par

\begin{figure}
    \centering
    \includegraphics[width=0.9\columnwidth]{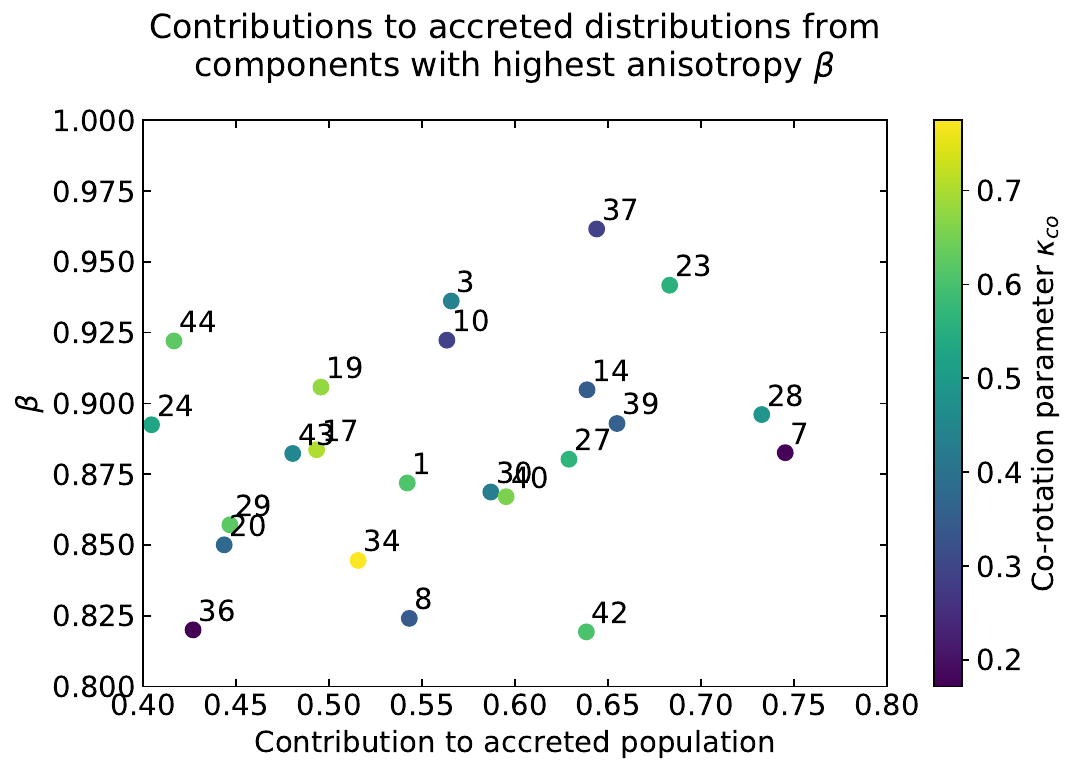}
    \caption{Selected region of Fig.~\ref{fig:beta_cont} containing galaxies with GS-like features (Sample~GS/all). The present-day co-rotation parameters $\kappa_\mathrm{co}$ are given by the colours of the points. Fifteen of the 23 galaxies shown have $\kappa_\mathrm{co}>0.4$ and are thus considered to be discs.}
    \label{fig:beta_cont_kappa}
\end{figure}

The selected region of Fig.~\ref{fig:beta_cont} is shown in Fig.~\ref{fig:beta_cont_kappa} colour-coded by $\kappa_\mathrm{co}$. Following \citet{Correa} we required that $\kappa_\mathrm{co}>0.4$ at the present-day for a galaxy to be considered a disc\footnote{Note however that $\kappa_\mathrm{co}$ is not a perfect indicator of the morphology of the galaxy \citep[see e.g.][]{Thob2019} and can slightly underestimate the fraction discs, depending on the threshold chosen}, although this threshold is somewhat arbitrary (see, e.g., \citealt{Font}). Fifteen of the 23 galaxies in Sample~GS/all survived this further cut. This set of fifteen galaxies is denoted Sample~GS/discs.\par
The co-rotation parameter can also be used to find what fraction of the \emph{disc} galaxies have a GS-like feature. Overall, 27 of the 45 \texttt{ARTEMIS} galaxies have $\kappa_\mathrm{co}>0.4$. Nine of these have both $\beta>0.8$ and a contribution of greater than $50\%$. Hence $1/3$ of the disc galaxies have a GS-like feature, the same fraction that was found without considering $\kappa_\mathrm{co}$.\par
\citet{Belokurov} and \citet{Fattahi} concluded that the GS-like features usually originate in a single merger with a massive satellite. To check this in the \texttt{ARTEMIS} galaxies we identified the most massive accreted progenitors (MMAPs) in each galaxy in Sample~GS/discs, by stellar mass. This was done by computing the peak stellar mass of any subhalo to which a star particle was bound before it was accreted onto the host. For two-thirds of the Sample~GS/discs galaxies, these MMAPs also had the highest dark matter (DM) and total masses of all accreted progenitors. It should be noted that due to stripping of the satellite prior to the merger, the peak stellar mass is reached before the final accretion. Hence some star particles are born after the peak and will not be registered as originating in the same satellite. However, since a large majority of stars are born before the peak in stellar mass, this is only a small fraction of the total number. The properties of these mergers will be summarised in Section \ref{section:properties}.\par

\begin{figure}
    \centering
    \includegraphics[width=\columnwidth]{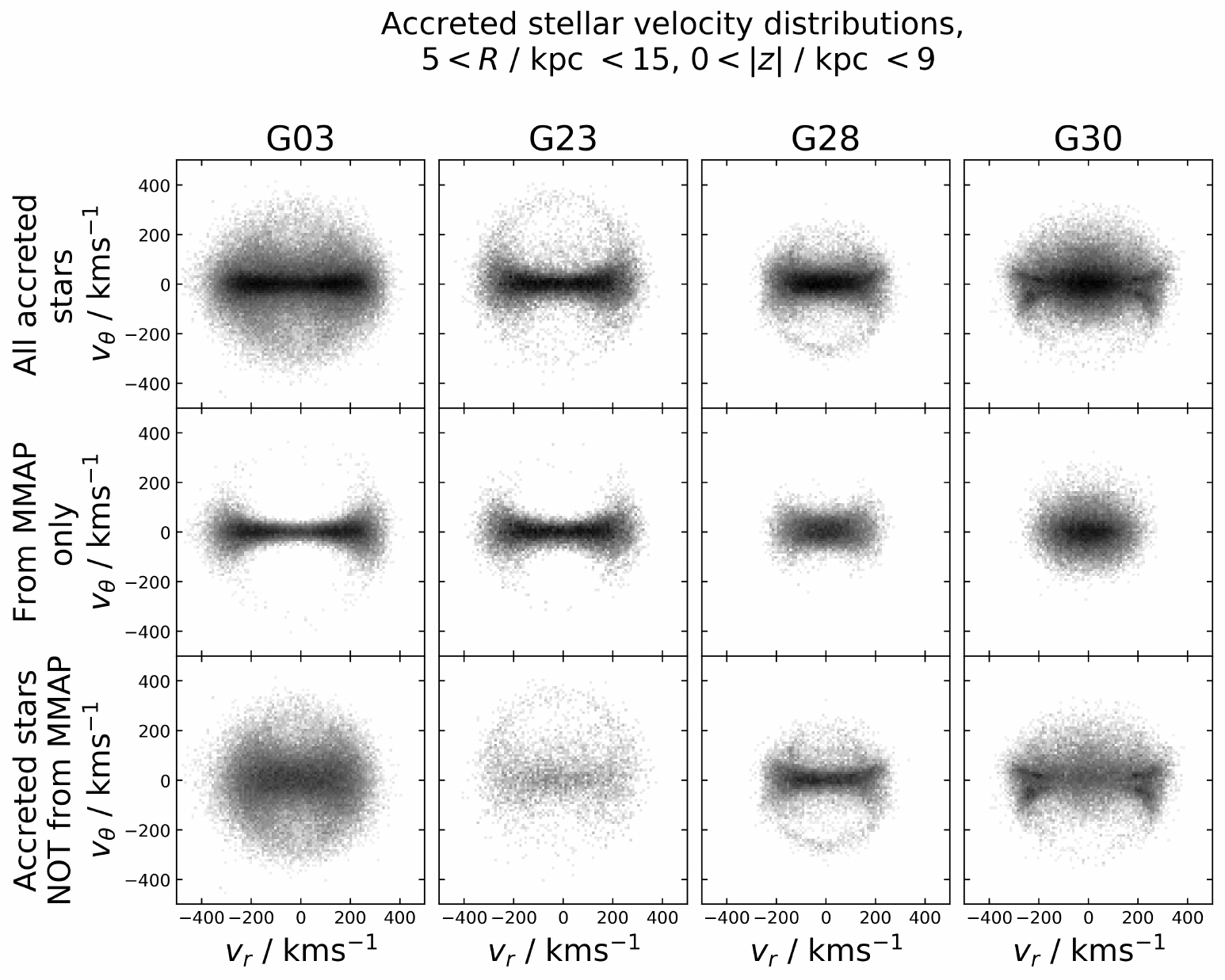}
    \caption{Spherical polar velocity distributions of all accreted stars (top row), stars from the MMAP (middle row) and accreted stars \emph{not} from the MMAP (bottom row), for four different Sample~GS/discs galaxies. For each galaxy (down each column) the greyscale normalisation is the same for each set of stars. Apart from in G30, the stars from the MMAPs visibly form the radially anisotropic components, with stars at higher $|v_\theta|$ removed. In G03 and G23 the stars from lower mass progenitors form more isotropic distributions, indicating that the GS-like feature is dominated by stars from the MMAP. In G28 a radially anisotropic feature is also visible in the lower plot, suggesting that the GS-like population is accreted in multiple mergers.}
    \label{fig:sausage_velocities}
\end{figure}
The velocity distributions of stars from the MMAPs are shown in Fig.~\ref{fig:sausage_velocities} for four different galaxies in Sample~GS/discs. For comparison, also shown are the distributions for all accreted stars, and for all accreted stars from other (lower mass) progenitors. The coordinate system and spatial cuts are the same as in Fig.~\ref{fig:fits}.\par

In G03 and G23, separating accreted stars according to whether they came from the MMAP effectively decomposes the velocity distribution into its constituent components. In both cases the stars from the MMAP form a highly radial feature with very few of the stars at high $|v_\theta|$. The distribution of other stars does not show such a feature clearly. This is in agreement with the conclusions of \citet{Belokurov} and \citet{Fattahi}. Most of the distributions of the Sample~GS/discs galaxies have a similar form. In G28 the radial feature visible in the top row has contributions from stars originating in both the MMAP and lower mass progenitors. Hence there are cases where the GS-like features are not formed solely by the MMAP mergers. These are different to the Milky Way, in which the GS is almost certainly the result of a single dominant accretion event~\citep{Belokurov}.\par
In G30 the MMAP stars form a more isotropic distribution, with the stars at high $|v_r|$ coming mostly from lower mass progenitors. This is only the case in two of the Sample~GS/discs galaxies, the other being G42. This may be related to the fact that the mass ratios of the MMAP mergers in these are the largest of all the selected galaxies (see Fig.~\ref{fig:mass_ratios_merger_times} in Section~\ref{section:properties}).

\subsection{Formation histories of the hosts}
\label{section:properties}
\begin{figure}
    \centering
    \includegraphics[width=0.9\columnwidth]{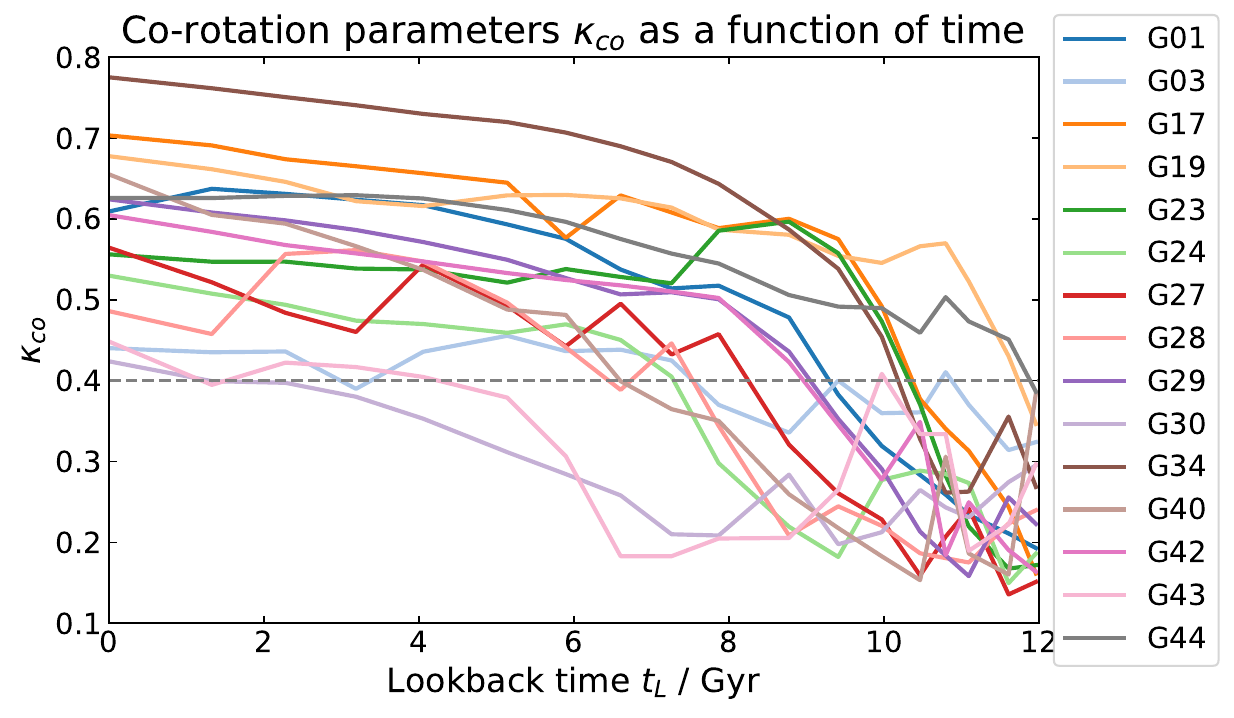}
    \caption{Co-rotation parameters $\kappa_\mathrm{co}$ of the Sample~GS/discs galaxies across time. The grey dashed line marking $\kappa_\mathrm{co}=0.4$ is the threshold above which a galaxy is considered to be a disc. As expected, $\kappa_\mathrm{co}$ usually increases towards the present day as stars are born in the disc. The different galaxies span a wide range of behaviours: for example G19 and G44 were already discs almost 12 Gyr ago, while G30 only reached $\kappa_\mathrm{co}=0.4$ in the last 3 Gyr.}
    \label{fig:kappa_co_t_L}
    \centering
    \includegraphics[width=0.9\columnwidth]{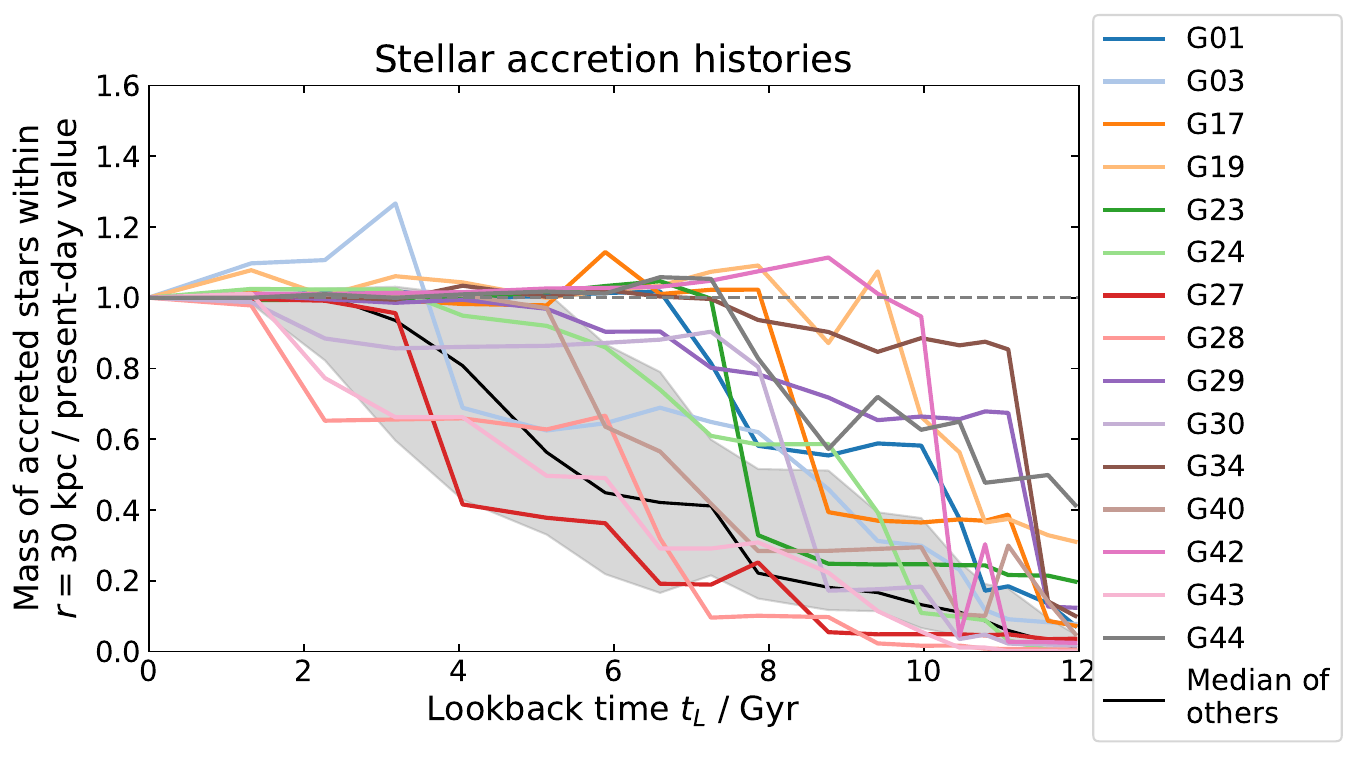}
    \caption{Total mass of accreted star particles within $r=30$~kpc, normalised by the present-day values. This effectively shows the galaxies' stellar accretion histories. The median for the galaxies not in Sample~GS/discs is plotted in black for comparison, with the interquartile range shown in grey. Ten of the fifteen Sample~GS/discs galaxies have already reached more than $90\%$ of their present-day accreted mass by $5$~Gyr ago. These have undergone no significant mergers in the last few billion years, similar to the Milky Way. By contrast, the non-Sample~GS/discs galaxies on average gain most of their mass later and more steadily.}
    \label{fig:accreted_mass_t_L}
\end{figure}

The galaxies in Sample~GS/discs are all discs at the present-day, with $\kappa_\mathrm{co}>0.4$. To quantify how the discs change across cosmic time, we calculated the values of $\kappa_\mathrm{co}$ (see eq.~\ref{equation:kappa}) at earlier times for each galaxy, using higher redshift snapshots of the simulations. The axis of rotation \hbox{($z$-axis)} was recalculated for each snapshot to account for any changes in orientation, again using the total angular momentum of star particles within $r=10$~kpc. The results are shown for all fifteen galaxies in Fig.~\ref{fig:kappa_co_t_L}, from a lookback time of \hbox{$t_L=12$ Gyr} up to the present-day.\par
In order to illustrate the galaxies' accretion histories, we calculated the total mass of accreted star particles within $r=30$ kpc as a function of time. These are shown in Fig.~\ref{fig:accreted_mass_t_L}. To allow comparison between the galaxies, the masses are all normalised to unity at the present-day. The median and interquartile range for the non-Sample~GS/discs galaxies are also plotted for comparison.

Figs.~\ref{fig:kappa_co_t_L} and \ref{fig:accreted_mass_t_L} demonstrate that the galaxies in Sample~GS/discs follow a broad range of evolutionary paths (this is expected in the cosmological context for disc galaxies, see e.g., \citealt{font2017}). The times at which the galaxies cross $\kappa_\mathrm{co}=0.4$ vary from more than 11~Gyr ago (G19 and G44) to less than 2 Gyr ago (G30 and G43). Similarly, the galaxies reach close to their present-day accreted masses at a range of times. G42 accreted very few stars in the last 10~Gyr, while as recently as 4~Gyr ago G27 had gained less than half of its accreted stellar mass.\par
However, all but two had gained at least $\sim60\%$ of their accreted stellar mass by $t_L=5$~Gyr. This is in contrast to typical non-Sample~GS/discs galaxies, which tend to accrete their mass more steadily and more recently. The Sample~GS/discs galaxies which accrete most of their mass early are comparable to the Milky Way. \cite{Naidu_substructure} showed that the stellar halo within $r=25$~kpc is dominated by the GS debris accreted over 6 Gyr ago. This is consistent with \citet{Deason_break}, who found that the broken radial density profile of the Milky Way's stellar halo is likely due to an early massive accretion event. \citet{Iorio} studied the shape of the Milky Way's stellar halo in more detail, and agreed that the majority of halo stars were deposited in such a merger event. As the figure shows the formation of the entire stellar halo one needs to take into account other important accretion events in the MW's life. For example, there is also the Sagittarius Dwarf Galaxy (Sgr) currently undergoing disruption \citep{Ibata}. This is estimated by \citet{Gibbons} to have had an initial dark halo mass of $\gtrsim 6\times 10^{10} {\rm M}_\odot$. However, a significant fraction of its stellar mass remains bound to the Sgr dwarf \citep[see e.g.,][]{Law, Vasiliev}, and \citet{Naidu_substructure} showed that the accreted Sgr stars only dominate the stellar halo beyond $r=25$~kpc.

\begin{figure}
    \centering
    \includegraphics[width=\columnwidth]{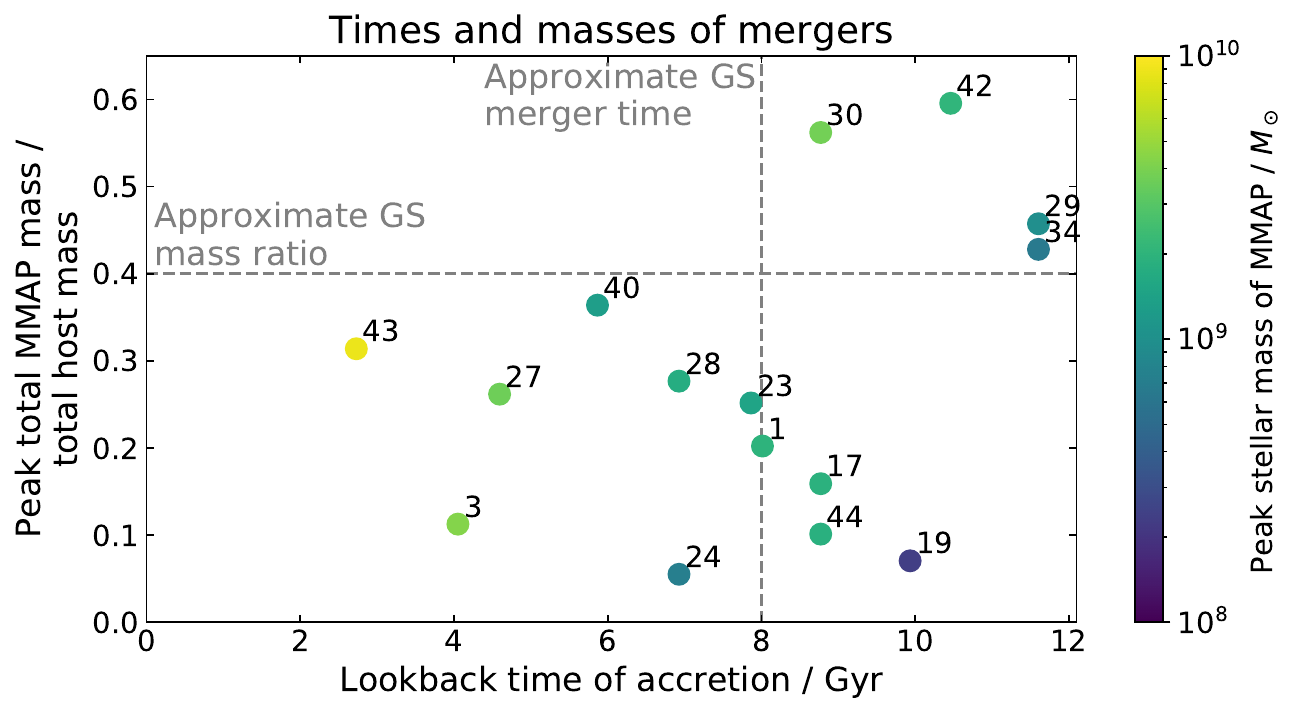}
    \caption{Merger times and mass ratios of the most massive merger events in Sample~GS/discs galaxies, with stellar masses colour-coded. The mass ratio is the peak total mass of the MMAP subhalo as a fraction of the total mass of the host at the same time. The grey dashed lines mark the approximate GS merger time of 8 Gyr ago, and a mass ratio of 0.4. This is the value estimated by \citet{Naidu_reconstruction} for the GS-Milky Way merger. The mergers span a wide range of times between $2$ and $12$ Gyr ago, although the majority are between $6$ and $10$ Gyr ago. There is also a wide range of mass ratios, between $\sim0.05$ and $\sim0.6$. The mergers with mass ratios greater than 0.4 all occur more than 8 Gyr before the present-day. In nearly all galaxies the peak stellar mass is on the order of $10^9\,{\rm M}_\odot$.}
    \label{fig:mass_ratios_merger_times}
\end{figure}

It is perhaps more instructive to focus on the merger time and the mass of the MMAP. In Fig.~\ref{fig:mass_ratios_merger_times} the accretion mass ratios and merger times of the MMAPs are plotted, colour-coded by peak stellar mass of the MMAP. We calculated the merger times from the 5th and 95th percentiles of redshifts at which stars were unbound from the MMAP and accreted onto the host. The plotted times are interpolations between these two values. The accretion redshift values are correct only to the nearest snapshot, which are separated by $\sim1$ Gyr or less; hence the accretion times are correct to within $\sim1$~Gyr. On the $y$-axis is the ratio of peak \emph{total} mass of the MMAP subhalo to the total mass of the host subhalo (measured at the same time). This may be of more relevance to this work than the stellar mass ratio alone, since any dynamical transformations of the host are likely to be affected by the total mass.

As expected from the accretion histories in Fig.~\ref{fig:accreted_mass_t_L}, there is a wide range of MMAP merger times. More than half of these are in the range 6-10 Gyr ago, broadly similar to the GS \citep[see][]{Belokurov,Helmi,splash,Dimatteo2019,Gallart2019,Bonaca2020}. The mass ratios span an order of magnitude between $\sim0.05$ and $\sim0.6$. This range contains the estimate of 2.5:1 (grey dashed line) by \citet{Naidu_reconstruction} for the corresponding mass ratio of the GS-Milky Way merger. However, we emphasise that the total mass ratio of the GS-Milky Way merger remains poorly constrained. This is because the total mass budget includes the dark matter and gas content of the dwarf, neither of which is observationally accessible. We therefore treat the total mass ratio as one of the least well constrained parameters of the interaction and do not wish to over-interpret it. Hence we do not limit our sample using this quantity. Mass ratios above 0.4 correspond to relatively early mergers (more than 8 Gyr ago). The only two cases with mass ratios greater than 0.5 (G30 and G42) are those whose MMAP stars appear to form more isotropic distributions at the present-day (see Fig.~\ref{fig:sausage_velocities}). While these mergers are somewhat different from the GS-Milky Way merger, we retain them in Sample~GS/discs since many of the resultant host transformations may be similar.\par
Almost all of the MMAPs have peak stellar masses on the order of $10^9\,{\rm M}_\odot$, with the only significant deviations being G19 (stellar mass $M_*\approx2\times10^8\,{\rm M}_\odot$) and G43 ($M_*\approx9\times10^9\,{\rm M}_\odot$). The others are all within the range $(5-50)\times10^8\,{\rm M}_\odot$ estimated by \citet{Myeong} for the GS. Other estimates for the stellar mass of the GS are similar, including $(5-10)\times10^8\, {\rm M}_\odot$ \citep[][]{Deason_mass}, $(7-70)\times10^8\, {\rm M}_\odot$ \citep[][]{Feuillet} and $\sim3\times10^8\,{\rm M}_\odot$ \citep[][]{Mackereth}.\par
Most of the galaxies in Sample~GS/discs are therefore comparable to the Milky Way, with accreted satellites whose properties are broadly consistent with our knowledge of the GS progenitor. We argue however that even the galaxies with later or earlier mergers can give insight into the processes which may have occurred in the Milky Way.

\section{Host transformations}
\label{sec:host_transformations}

As described in Section~\ref{section:introduction}, previous studies have revealed a range of possible effects that mergers may have on the host galaxies. In this section we investigate two such transformations in \texttt{ARTEMIS}, namely disc flipping and disc splashing. But firstly we look at a larger-scale effect: changes to the masses and shapes of the DM haloes surrounding the galaxies. For these purposes we use the fifteen galaxies in Sample~GS/discs described in Section~\ref{section:properties}. 

\subsection{Dark matter halo evolution}
\label{section:DM}
Before studying the details of the evolution of the DM halo shapes, it is worth pointing out that DM halos behave very differently in the absence (DM-only simulations) and in the presence of baryons. In particular, the DM radial density profiles can be modified when gas piles up at the bottom of the potential well or is rapidly removed from the galaxy's centre due to feedback processes \citep[e.g.][]{Blumenthal1986,Gnedin2004,Mashchenko2006,Pontzen2012,Dicintio2014}. Formation of a dense baryonic disc in the centre of the halo forces the DM to adjust by reshaping and aligning. As a result, the inner portions of DM halos that host a disc are much more spherical than those that do not \citep[see e.g.][]{Dubinski1994,Abadi2010}. The sculpting of the DM shape is a complex process: the DM halo can retain some memory of the initial collapse set by the surrounding large scale structure \citep[see][]{VeraCiro2011}, its central regions can be molded by the baryons that settle there, while its outskirts reflect the most recent accretion events \citep[see e.g.][]{Shao2021}.\par
To quantify the accretion histories of the DM haloes, we calculated the total masses of DM within the radial ranges $r<30$~kpc and $30$~kpc~$<r<100$~kpc. This is analogous to the calculations of stellar mass used for Fig.~\ref{fig:accreted_mass_t_L}.\par
\begin{figure*}
    \centering
    \includegraphics[width=\textwidth]{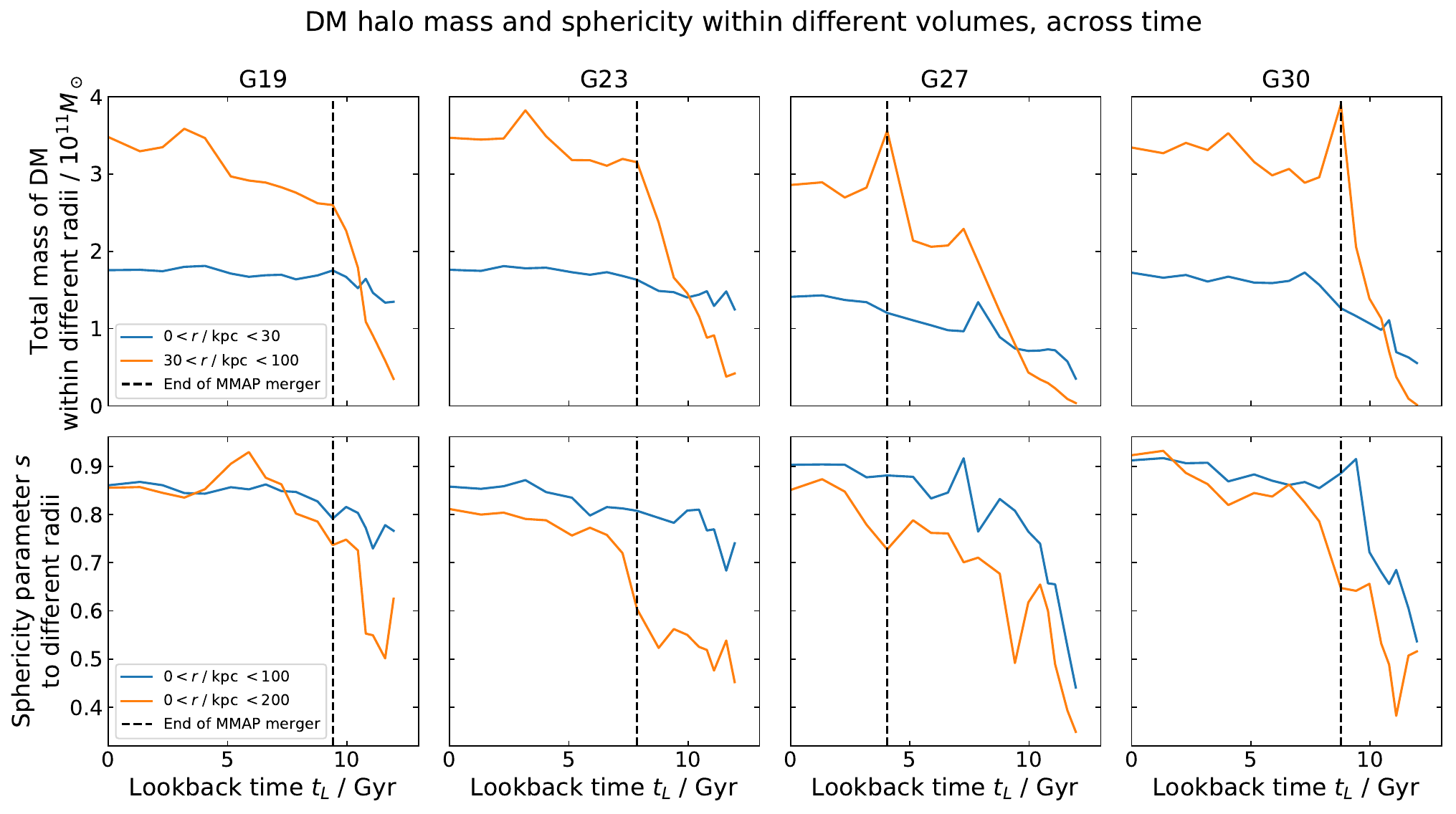}
    \caption{\textbf{Top row:} total masses of DM between radii of 0 and 30~kpc (blue) and between 30 and 100~kpc (orange), against lookback time, in four different galaxies in Sample~GS/discs. In each case large masses of dark matter are gained in the outer regions, especially prior to the MMAP merger (black dashed line). The mass inside 30~kpc increases by a much smaller amount. \textbf{Bottom row:}~evolution of sphericities $s$ of the DM haloes, measured using the positions of particles within radii of 30~kpc (blue) and 100~kpc (orange). In both cases $s$ increases across time in each galaxy, indicating that the haloes become more spherical. In some galaxies the MMAP mergers roughly coincide with rapid increases in $s$ (e.g., in G23 and G30).}
    \label{fig:DM_halo}
\end{figure*}

A simple commonly used measure of the shape of a DM halo is its sphericity $s$, calculated as follows. Using essentially the same method as \citet{Springel}, we found the principal axes of the halo. Consider the tensor
\begin{equation}
    I_{jk}=\sum_{\substack{\mathrm{DM}\\\mathrm{particles}}}^{r<30\mathrm{kpc}}x_jx_k,
\end{equation}
 where $x_j$ are the Cartesian coordinate components of DM particles in the frame centred on the host's CoP.
The sum is over all DM particles within a radius of $r=30$~kpc. The minor axis of the halo within this radius is given by the eigenvector corresponding to the smallest eigenvalue, and the major axis by the eigenvector corresponding to the largest eigenvalue. Let the (positive) eigenvalues of $I_{jk}$ be $a^2$, $b^2$ and $c^2$, with $a^2\geq b^2\geq c^2$. The values $a$, $b$ and $c$ are then proportional to the root mean square distances of these particles from the origin, along each of the principal axes. We define the sphericity as $s=c/a$, such that $s=1$ if the halo is spherical, and $s=0$ if it is razor-thin. This was calculated for all fifteen galaxies in Sample~GS/discs across many snapshots. It was also repeated for all DM particles within a radius of $r=100$~kpc. These masses and sphericity parameters are plotted for four different Sample~GS/discs galaxies across time in Fig.~\ref{fig:DM_halo}.\par

In each of the four galaxies shown, more than $10^{11}\, {\rm M}_\odot$ of DM is added to the region between radii 30~kpc and 100~kpc in the few billion years prior to the MMAP merger, after which far less mass is gained. However, the mass within $r=30$~kpc remains roughly constant throughout this accretion of matter to the outer halo. Most of the Sample~GS/discs galaxies show similar behaviour. The merger with the MMAP galaxy is therefore a rough marker of the end of DM accretion to the halo, at least within a radius of 100~kpc. The bottom row of Fig.~\ref{fig:DM_halo} demonstrates that the shapes of the haloes also evolve. In each of the four galaxies, and in virtually all of the galaxies in Sample~GS/discs, the sphericity increases with time. The most rapid increases in $s$ for particles within 100~kpc often coincide with the periods of fastest accretion to the outer parts of the halo (e.g., in G19, G23 and G30). The value of $s$ remains more stable when little DM is being accreted, for example in the last 6~Gyr in G23. It therefore appears that accretion of DM to the halo is associated with the shape of the halo becoming more spherical.\par
To quantify the number of Sample~GS/discs galaxies which undergo a large sphericity change within $r<100$~kpc, we proceeded as follows. We linearly interpolated $s$ as a function of lookback time between each snapshot, and found the changes in $s$ over all 2~Gyr-long periods. 7 of these galaxies experience increases in $s$ of greater than 0.2 during a 2~Gyr period which includes the time of the MMAP merger. These include G19, G23 and G30, but not G27. By comparison, the mean change in $s$ across all 2~Gyr periods since $t_L=12$~Gyr is less than 0.1 in all 15 galaxies.  Hence by this definition, almost half of the Sample~GS/discs galaxies experience large increases of halo sphericity coincident with MMAP mergers.\par
Changes in orientation of the DM halo principal axes are discussed in relation to disc flipping in Section~\ref{section:disc_flipping}.

\subsection{Disc flipping}
\label{section:disc_flipping}
There is reason to believe \citep[e.g.,][]{Bett, Welker, Dekel} that the angular momentum (AM) axes of DM, stellar and gaseous components of Milky Way-like galaxies experience rapid orientation changes (flips) caused by mergers.
To show whether this is the case in the \texttt{ARTEMIS} galaxies, we calculated the net AM orientations across time of a) all \textit{in situ} star particles within $r=30$~kpc, and b) only those born before $t_L=10$ Gyr. Here we choose to exclude accreted stars from the calculation to ensure that any changes in orientation are not only due to the AM of newly accreted stars. We later found that this choice of definition did not affect our conclusions.\par
Using the method described in Section~\ref{section:DM}, we found the principal axes of the DM halo. In the Milky Way there is reason to believe that the DM halo minor axis direction changes between inner and outer radii \citep[e.g., see][]{Shao, Deason_mismatch}. Motivated by this, as in Section~\ref{section:DM} we calculated the tensor $I_{jk}$ twice, for DM particles within $r=30$~kpc and $r=100$~kpc.\par
\begin{figure*}
    \centering
    \includegraphics[width=0.9\textwidth]{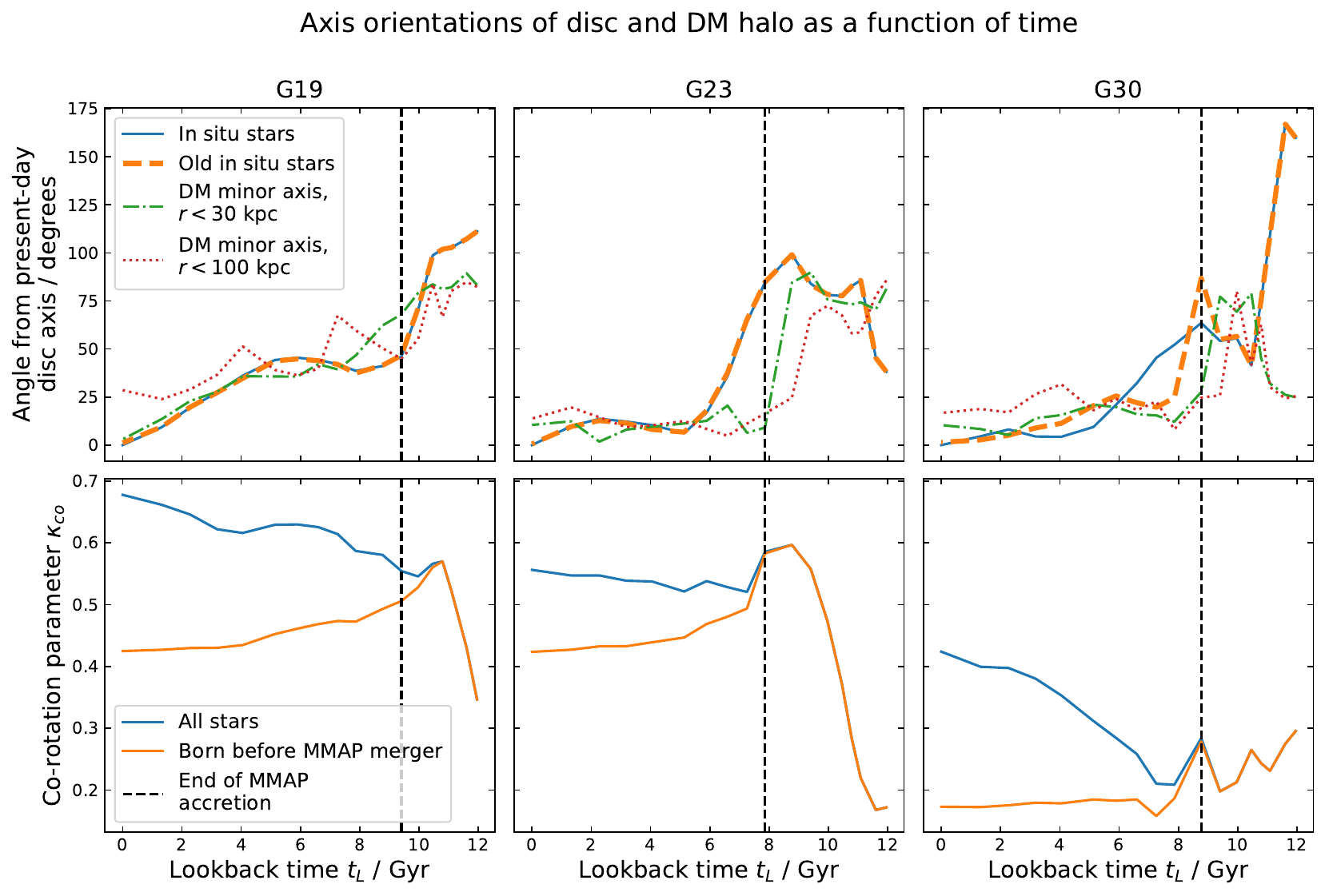}
    \caption{\textbf{Top row:} angles of \textit{in situ} AM axes and DM minor axes, all measured from the present-day \textit{in situ} axis. The `old \textit{in situ} stars' are born before $t_L=10$ Gyr. In G19 and G23 the orange and blue traces are very close, so are difficult to distinguish. In these two galaxies the AM direction changes by more than $\sim50^\circ$ in a period of less than $\sim2$ Gyr around the MMAP merger. The DM minor axes are reasonably well-aligned with the disc after this flip, particularly for $r<30$ kpc. In G30 the AM of old \textit{in situ} stars does experience a rapid change at the time of the merger, but it is not as closely coupled to the overall AM. \textbf{Bottom row:} $\kappa_\mathrm{co}$ measured for all stars within 30 kpc (blue) and those born before the beginning of stellar accretion from the MMAP (orange). In G19 and G23 $\kappa_\mathrm{co}$ for all stars remains above 0.4 during the flip, implying that the stellar disc survives the flip. In G30 $\kappa_\mathrm{co}$ is smaller than 0.4 at the time of the merger, so the AM change is not classed as a disc flip.}
    \label{fig:disc_angles}
\end{figure*}

The angles of the AM axes and the DM minor axes are plotted in Fig.~\ref{fig:disc_angles}, for some of the galaxies which \emph{do} experience AM orientation changes. All angles are measured relative to the same direction, namely that of the \textit{in situ} AM at the present-day.\par
For a spin flip to also be a \emph{disc} flip, the galaxy must be a disc at the time of flipping. For this reason it is instructive to also plot the co-rotation parameter $\kappa_\mathrm{co}$ (see eq.~\ref{equation:kappa}). This is calculated for two sets of stars: a) all stars within $r=30$ kpc, and b) stars within $r=30$~kpc born \emph{before} the start of MMAP accretion. This will show a) whether a disc can survive a flip, and b) whether stars born before the flip remain in a disc configuration themselves. In both cases, as always, the $z$-axis of the coordinate system used to define $\kappa_\mathrm{co}$ is aligned with the total angular momentum of stars within $r=10$~kpc. These results are also plotted \hbox{in Fig.~\ref{fig:disc_angles}.}

The most striking features of the top row of Fig.~\ref{fig:disc_angles} are rapid changes in the orientations of the disc AM at around the MMAP merger times. In G19 and G23 the AM axis of \textit{in situ} stars changes orientation by more than $\sim50^\circ$ in less than 2 Gyr. In G23 the flip is more than $90^\circ$. The dashed orange traces show that this is not merely the result of new stars being born in a new disc with a different orientation: these are all stars born \emph{before} the AM flip. The AM direction of these stars is almost indistinguishable from that of the whole \textit{in situ} population. We have checked that these results are largely unchanged when the axis of the disc is defined differently (e.g., including accreted stars, calculating AM within different radii, or using the minor axis of an inertia tensor).\par
Furthermore, the bottom row of plots shows that the discs of G19 and G23 remain discs throughout their flips. $\kappa_\mathrm{co}$ calculated for all stars remains comfortably above $0.4$ during the time of fastest flipping. This remains true when $\kappa_\mathrm{co}$ is calculated for only stars born before the merger. This shows that old stars are remaining in a disc configuration even when the axis of that disc is changing by large angles. The decreases in $\kappa_\mathrm{co}$ seen in each case are likely related to the phenomenon of splashing (see Section \ref{section:splashing}).\par
However, the changes in AM orientation are not limited to the time of the merger. In many cases, the AM axis undergoes slower precession even when the galaxy is not accreting stars at a significant rate. In Fig.~\ref{fig:disc_angles} this is seen most clearly in G19, where the axis direction steadily changes by almost $50^\circ$ in 6 Gyr. It can be seen from Fig.~\ref{fig:accreted_mass_t_L} that this galaxy does not accrete any significant mass of stars in this time. A similar effect in other simulations was described by \citet{McCarthy}, who found that star-forming discs are torqued across cosmic time. They showed that this is so severe that there is almost no correlation between the disc orientations at redshift $z=2$ ($t_L\sim10$ Gyr) and the present-day. The modern disc orientation is therefore the result of these two apparently different effects: rapid disc flipping and slow disc precession.\par
G30 exhibits contrasting behaviour. While the AM of old \textit{in situ} stars rapidly changes orientation at the time of the MMAP merger, its direction is not coupled strongly the AM of all \textit{in situ} stars, which precesses at a different rate. Furthermore, the value of $\kappa_\mathrm{co}$ is less than 0.3 at the merger time. For the stars born before this time, the value decreases to below 0.2 until the present-day. Hence even though the AM changes orientation, we do not consider G30 to have undergone a \emph{disc} flip. While there also appears to be a large flip at 12 Gyr ago, we note that the number of star particles in the halo at this time is very small. Hence this change does not represent a major transformation to the galaxy.\par

In G19 and G23, the DM minor axis within $r=30$ kpc (green dash-dot curve) appears well-aligned with the axis of the disc (solid blue curve) from soon after the rapid flip onward. This is indeed the case: the angle between them is almost always less than $10^\circ$ in these time periods. There is similarly close alignment in G30 with the old \textit{in situ} stars. This is consistent with the inferences of \citet{Shao} that the minor axis of the Milky Way's inner DM halo is perpendicular to its disc. Other work on simulations \citep[e.g.,][]{Hahn} also shows that the minor axes of inner DM haloes tend to be well-aligned with stellar and gas discs \citep[see also][]{Debattista2008,Debattista2015}.\par
It is notable that in both G23 and G30 the stellar AM flip is preceded by a flip in the DM 30~kpc minor \hbox{axis $\sim1$ Gyr} earlier, which is in turn preceded by a flip in the DM 100~kpc minor axis. This hints that the rapid re-orientation might first take place at large radii, with shells of the DM halo at smaller radii flipping later, followed finally by the disc. However, this is not seen as clearly in G19.\par
Note that while the fastest flips coincide roughly with the MMAP mergers, this does not necessarily imply that these mergers are directly responsible for the flips. For example, large masses of DM not bound to the MMAP may be accreted at a similar time; e.g., \citet{Genel} found in simulations that around 40\% of DM halo mass originates from smooth accretion of previously unbound DM. If the mass of this matter is significant, it may well be partially responsible for causing the angular momentum of the host to change.\par
G30 is one of the four galaxies which experienced early, high mass ratio mergers. The other three all show similar $\kappa_\mathrm{co}$ evolution, but in G29 and G34 the AM of the two sets of \textit{in situ} stars are more closely coupled. Both G29 and G42 exhibit rapid changes in their stellar AM orientation, at the same times as their MMAP mergers. Hence rapid AM flips occur in galaxies both with and without discs, and in cases with both small and large merger mass ratios.

\subsubsection{Number of disc flips}
To find the number of galaxies in Sample~GS/discs which experience a disc flip, we proceeded as follows. In order to estimate the angular momentum orientations as continuous functions of time, we interpolated the AM unit vectors between the snapshots, using the approximation that the vectors rotated in fixed planes with constant angular speed between consecutive snapshots.\par
The definition of a disc flip is highly arbitrary, made more so by the fact that the AM orientations precess slowly even when not flipping rapidly. We adopt the definition that a disc flip is a change of \textit{in situ} AM orientation by more than $60^\circ$ in any period of 2~Gyr, during which $\kappa_\mathrm{co}>0.4$ throughout. The period of 2~Gyr is chosen because it roughly corresponds to the duration of the large AM swing clearly seen in G23. With this definition, the number of galaxies which have undergone disc flips is four: namely G19, G23, G24 and G27. This result is unchanged if we instead require angle changes of $45^\circ$ in 2~Gyr, or changes of $30^\circ$ in 1~Gyr. Note that G30 is excluded because $\kappa_\mathrm{co}<0.4$ during its large AM changes. In all four of these galaxies, the flips coincide with the MMAP mergers, supporting the hypothesis that these events are linked. There are no galaxies which experience two or more disc flips, according to these definitions. This may simply be because disc flipping events are sufficiently rare that we do not see any repeats in our sample. If only 4/15 of the galaxies in the sample experience a single disc flip, then it is reasonable that none of these experience a second one.\par

\subsubsection{Maximum rates of flipping}
In addition to finding the presence of disc flips, We also calculated the maximum rates of change of the disc axis directions. Since most discs appear to undergo precession at all times, even when not accreting, this is less arbitrary than an absolute change in orientation. Let $\hat{\bm\omega}_i$ be the unit vector aligned with the AM vector of all \textit{in situ} stars within \hbox{$r=30$ kpc}, in the simulation snapshot $i$. Then the change in AM orientation $\Delta\delta_i$ between snapshots $i$ and $i+1$ is given by $\mathrm{cos}\:\Delta\delta_i=\hat{\bm\omega}_i\cdot\hat{\bm\omega}_{i+1}$. Also let $\Delta t_i=t_{i+1}-t_{i}$ be the time interval between these snapshots, where $t_i$ is the cosmic time at snapshot $i$ ($t_i$ increases towards the present-day). The mean rate of change of the AM orientation between these two snapshots is then
\begin{equation}
    \Omega_i=\frac{\Delta\delta_i}{\Delta t_i},
\end{equation}
in units of radians Gyr$^{-1}$. We repeated this calculation for all snapshots in the range $i_{\mathrm{start}}-1\leq i\leq i_{\mathrm{end}}+1$ where $i_{\mathrm{start}}$ is the snapshot at the approximate beginning of stellar accretion of the MMAP, and $i_{\mathrm{end}}$ is at the end of accretion. This corresponds to a few Gyr around the time of the merger. To obtain an estimate of the maximum precession rate associated with the merger, we calculated the maximum value of $\Omega_i$ across these intervals. Due to the discrete nature of the snapshots this is not an instantaneous maximum rate, but a maximum averaged rate over periods of \hbox{$\sim1$ Gyr.} Note that this may be somewhat sensitive to the temporal resolution of the snapshots. In particular, if the flips happen in time intervals shorter than the gaps between snapshots, the maximum rates would be underestimated. However, it can be seen in Fig.~\ref{fig:disc_angles} that the large changes happen over 3-4 snapshots. This is usually the case in the other galaxies. This method is therefore adequate for providing estimates of the maximum rates.\par

\begin{figure}
    \centering
    \includegraphics[width=\columnwidth]{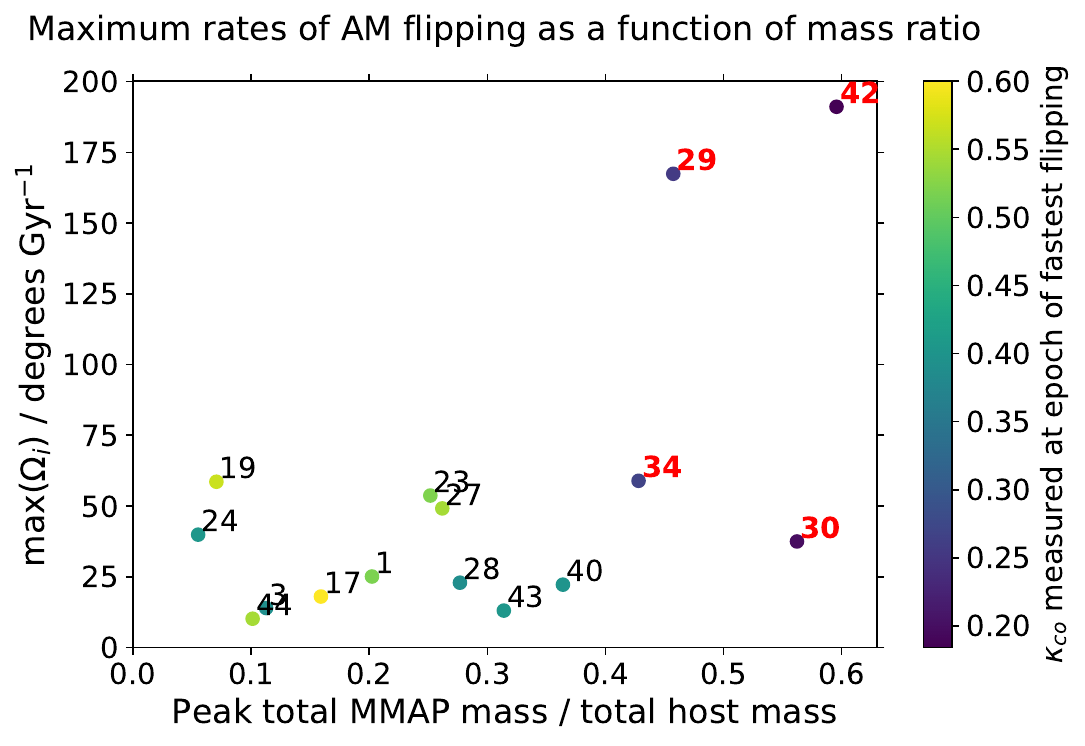}
    \caption{Maximum rates of AM flipping associated with mergers against total mass ratio for Sample~GS/discs galaxies. The mass ratio is the same as that in Fig.~\ref{fig:mass_ratios_merger_times}. The co-rotation parameter $\kappa_\mathrm{co}$ is colour-coded. Extremely fast rates (more than 100 degrees Gyr$^{-1}$) are only seen where the mass ratio is large (greater than 0.4), and $\kappa_\mathrm{co}$ is small. In cases where there is a disc ($\kappa_\mathrm{co}\gtrsim0.4$), the fastest rates do not exceed $\sim60$ degrees Gyr$^{-1}$. Below a mass ratio of 0.4 there is no apparent correlation between mass ratio and peak flipping rate.}
    \label{fig:angle_rates}
\end{figure}

The maximum rates $\mathrm{max}(\Omega_i)$ are plotted in Fig.~\ref{fig:angle_rates} against the total mass ratio, defined as in Fig.~\ref{fig:mass_ratios_merger_times}. The points are colour-coded by the co-rotation parameter $\kappa_\mathrm{co}$, measured at the snapshot corresponding to $\mathrm{max}(\Omega_i)$. This shows whether the stars do indeed form a disc (i.e., $\kappa_\mathrm{co}>0.4$) when the AM is flipping at its fastest rate. If $\kappa_\mathrm{co}$ is small the flipping rate has less meaning, since less of the stellar motion is associated with the net AM.\par

In galaxies where $\kappa_\mathrm{co}>0.4$, Fig.~\ref{fig:angle_rates} shows that $\mathrm{max}(\Omega_i)$ takes values between 10 and 60 degrees Gyr$^{-1}$. Note that the four galaxies which experience a disc flip according to our definition (G19, G23, G24 and G27) exhibit the fastest rates of all the galaxies with $\kappa_\mathrm{co}>0.4$. The rate only exceeds 60 degrees Gyr$^{-1}$ in G29 and G42, where $\kappa_\mathrm{co}<0.3$. It can be seen from Fig.~\ref{fig:mass_ratios_merger_times} that the MMAP mergers in both of these galaxies occur early ($t_L>10$~Gyr), with a high mass ratio. Fig.~\ref{fig:kappa_co_t_L} also shows that $\kappa_\mathrm{co}$ is small for these galaxies before their respective mergers, as well as during and immediately after. Since a collection of stars with small $\kappa_\mathrm{co}$ is not experiencing coherent rotation to the same degree as a disc, rapid changes of angular momentum direction are less significant events. 

Therefore, we conclude that the highest observed \emph{disc} flipping rates coincident with mergers do not exceed 60 degrees Gyr$^{-1}$. For a particle with an orbital period of $500$ Myr (a realistic value for a Milky Way-like galaxy), this corresponds to a rate of 30 degrees per orbit. This is consistent with \citet{Dekel}, who predict using simulations that gas discs undergo and survive flips (defined as changes of more than $45^\circ$) in less than an orbital period.\par
There is no obvious correlation between total mass ratio of the merger and maximum flipping rate. The disc of G19 flips at a maximum rate \hbox{of $\sim60$ degrees Gyr$^{-1}$} despite a small total mass ratio of $\sim0.05$. There are many reasons why this could be the case, such as merger geometry. It must also be remembered that the total mass ratio only takes account of particles bound to the MMAP's \emph{subhalo}, not any associated satellites or companions. In reality, the actual mass being accreted around the time of the merger may be much greater than merely that of the MMAP. While Fig.~\ref{fig:disc_angles} proves that the disc flips are contemporaneous with the MMAP events, the lack of correlation shown in Fig.~\ref{fig:angle_rates} could be viewed as a demonstration that the flipping is not caused by the MMAP itself. Note that in some cases it is also possible that the flipping is due to causes other than the mergers. \citet{Bett} found in simulations of Milky Way-mass haloes that large spin flips of inner DM haloes frequently occur in the absence of mergers. For example, these could be caused by tidal interactions during flybys of passing objects. The spin flipping histories of galaxies are therefore complex and varied. However, since a significant merger did occur in the Milky Way's past, it is entirely possible that it caused the Milky Way to experience a significant flip.

\begin{figure*}
    \centering
    \includegraphics[width=0.8\textwidth]{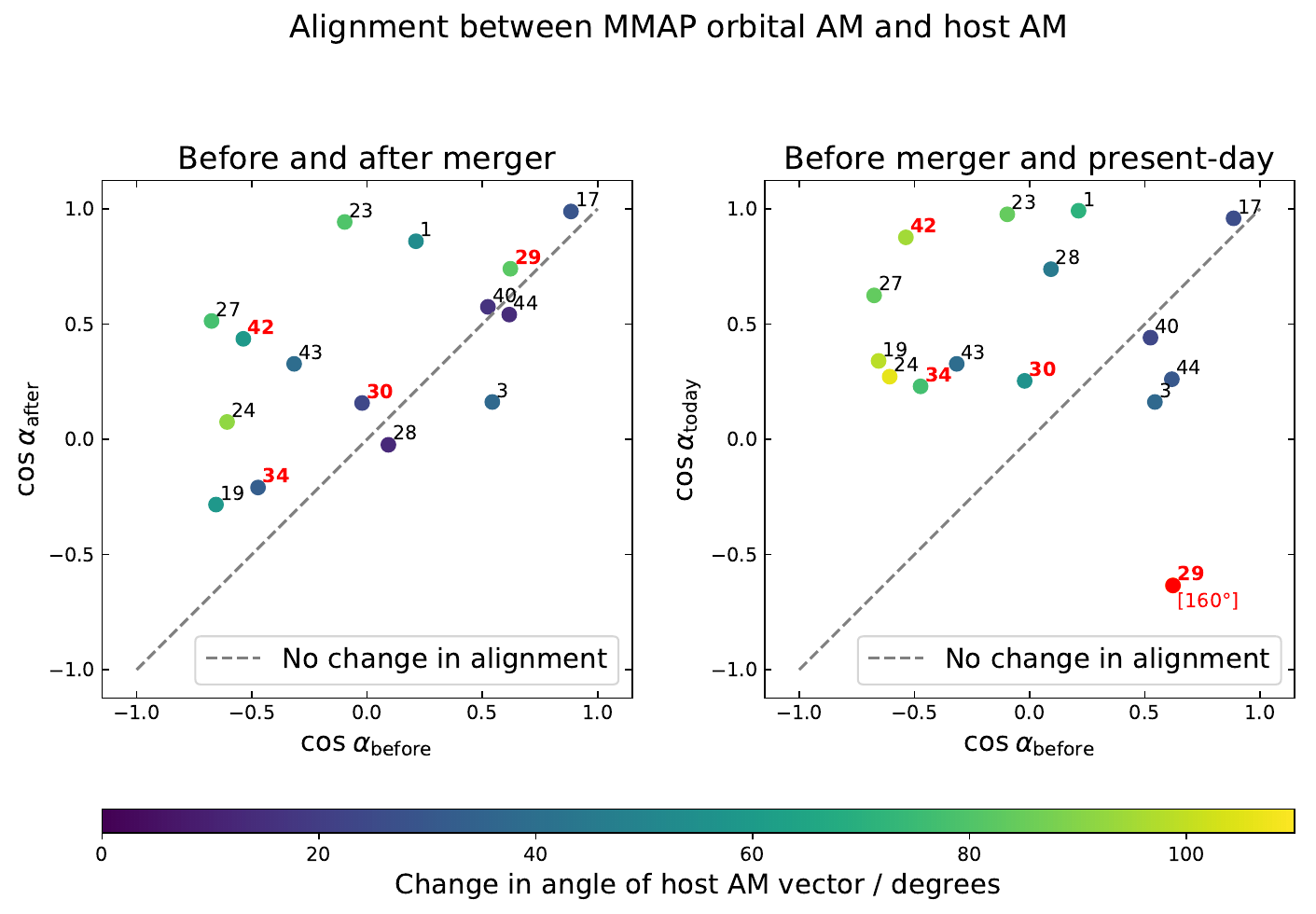}
    \caption{\textbf{Left-hand panel:} cosines of the angles between the MMAP OAM and the host AM three snapshots before ($\alpha_\mathrm{before}$) and after ($\alpha_\mathrm{after}$) the merger, for all Sample~GS/discs galaxies. \textbf{Right-hand panel:} the same, but the host AM angle after the merger is measured at the present-day ($\alpha_\mathrm{today}$). The MMAP OAM is measured three snapshots before the merger in all cases. The points are colour-coded by the change in angle of the host AM vector between the two relevant snapshots. For G29 this value ($160^\circ$) is outside the range of the colourbar, so is given separately. Points on the grey dashed lines represent no change in relative alignment with the satellite's OAM. Unsurprisingly, points closest to these lines also correspond to galaxies whose AM vectors change by the smallest angles (i.e., shaded purple). Those further away from these lines experience a greater change in both angle and relative alignment with the satellite's OAM. In both panels most of the points lie above the grey dashed lines, corresponding to galaxies whose alignment with the pre-merger orbital plane increases. At the present day all but one of the galaxies have disc axes aligned within $90^\circ$ of the OAM. Note that in the right-hand panel G29 and G42 have experienced some of the most extreme changes. These are the galaxies with small $\kappa_\mathrm{co}$ which experience the most rapid angular momentum flips.}
    \label{fig:orbital_AM_alignment}
\end{figure*}

\subsubsection{Connection to merger geometry}
It might be expected that the change in AM direction of the host is related to the orbital angular momentum (OAM) of the satellite's orbit about the host prior to accretion. To test this, we used the merger trees to find the centre of mass (CoM) positions and peculiar velocities of the host and MMAP before the merger. We transformed from comoving to physical coordinates and into the reference frame centred on the host at each snapshot. This allowed the OAM direction of the satellite's orbit to be calculated (i.e., the normal to its orbital plane with respect to the host). Let $\bm{r}_i$ be the position vector of the satellite's CoM relative to the host's CoM in the $i$th snapshot and $\bm{v}_i$ its relative peculiar velocity. The unit normal vector to the orbital plane is then
\begin{equation}
    \hat{\bm{J}}_i=\frac{\bm{r}_i\times\bm{v}_i}{|\bm{r}_i\times\bm{v}_i|}.
\end{equation}\par
Also let $a$ be the index of the final snapshot in which the MMAP is present before accretion. To find the orbital plane we calculated $\hat{\bm{J}}_{a-2}$ (i.e., three snapshots before final accretion). This was chosen because the merger is more likely to be already underway in later snapshots. We also calculated the host's stellar AM directions $\hat{\bm\omega}_i$ in the same snapshot, in the snapshot $a+3$ (three snapshots after MMAP accretion), and at the present-day snapshot $p$. We later checked that the results were not sensitive to the exact snapshots chosen. To quantify the alignment between the orbital plane and the \textit{in situ} AM before and after the merger, define the angles $\alpha_\mathrm{before}$, $\alpha_\mathrm{after}$ and $\alpha_\mathrm{today}$ by
\begin{align}
    \mathrm{cos}\:\alpha_\mathrm{before}&=\hat{\bm{J}}_{a-2}\cdot\hat{\bm\omega}_{a-2}, \\
    \mathrm{cos}\:\alpha_\mathrm{after}&=\hat{\bm{J}}_{a-2}\cdot\hat{\bm\omega}_{a+3}, \\
    \mathrm{cos}\:\alpha_\mathrm{today}&=\hat{\bm{J}}_{a-2}\cdot\hat{\bm\omega}_{p}.
\end{align}
These are the angles of the AM axis three snapshots before and after the merger and at the present-day, all measured relative to the OAM axis of the MMAP before the merger. If \hbox{$\mathrm{cos}\:\alpha_\mathrm{after}>\mathrm{cos}\:\alpha_\mathrm{before}$}, the AM of the host is more aligned with the satellite's (pre-merger) OAM after the merger than it was before. To show whether this is the case, $\mathrm{cos}\:\alpha_\mathrm{after}$ is plotted against $\mathrm{cos}\:\alpha_\mathrm{before}$ in the left-hand panel of Fig.~\ref{fig:orbital_AM_alignment} for the fifteen galaxies. This is repeated with $\mathrm{cos}\:\alpha_\mathrm{today}$ instead of $\mathrm{cos}\:\alpha_\mathrm{after}$ in the right-hand panel.\par
Before the merger, Fig.~\ref{fig:orbital_AM_alignment} shows no clear preferred orientation of the host's AM with respect to the satellite's orbital plane. The alignment $\mathrm{cos}\:\alpha_\mathrm{before}$ takes a wide range of values, and has a median close to zero (as expected for two uncorrelated directions). However, 12/15 of the galaxies have $\mathrm{cos}\:\alpha_\mathrm{after}>0$. Hence in 80\% of cases the host's post-merger AM lies within $90^\circ$ of the MMAP's pre-merger OAM. More importantly, 12/15 of the points lie above the grey dashed line \hbox{$\mathrm{cos}\:\alpha_\mathrm{after}=\mathrm{cos}\:\alpha_\mathrm{before}$.} Points above this line have \hbox{$\mathrm{cos}\:\alpha_\mathrm{after}>\mathrm{cos}\:\alpha_\mathrm{before}$}, meaning that the host AM is more closely aligned with the satellite's OAM after the merger than before. An even stronger pattern can be seen in the right-hand panel of Fig.~\ref{fig:orbital_AM_alignment}. There is a clear preference for the present-day disc AM to be at least partially aligned with the satellite's OAM, with only one case (G29) where the directions differ by more than $90^\circ$. In addition to G29 there are only three other cases where the alignment was closer before the merger than at the present-day.\par

\begin{figure*}
    \centering
    \includegraphics[width=0.9\textwidth]{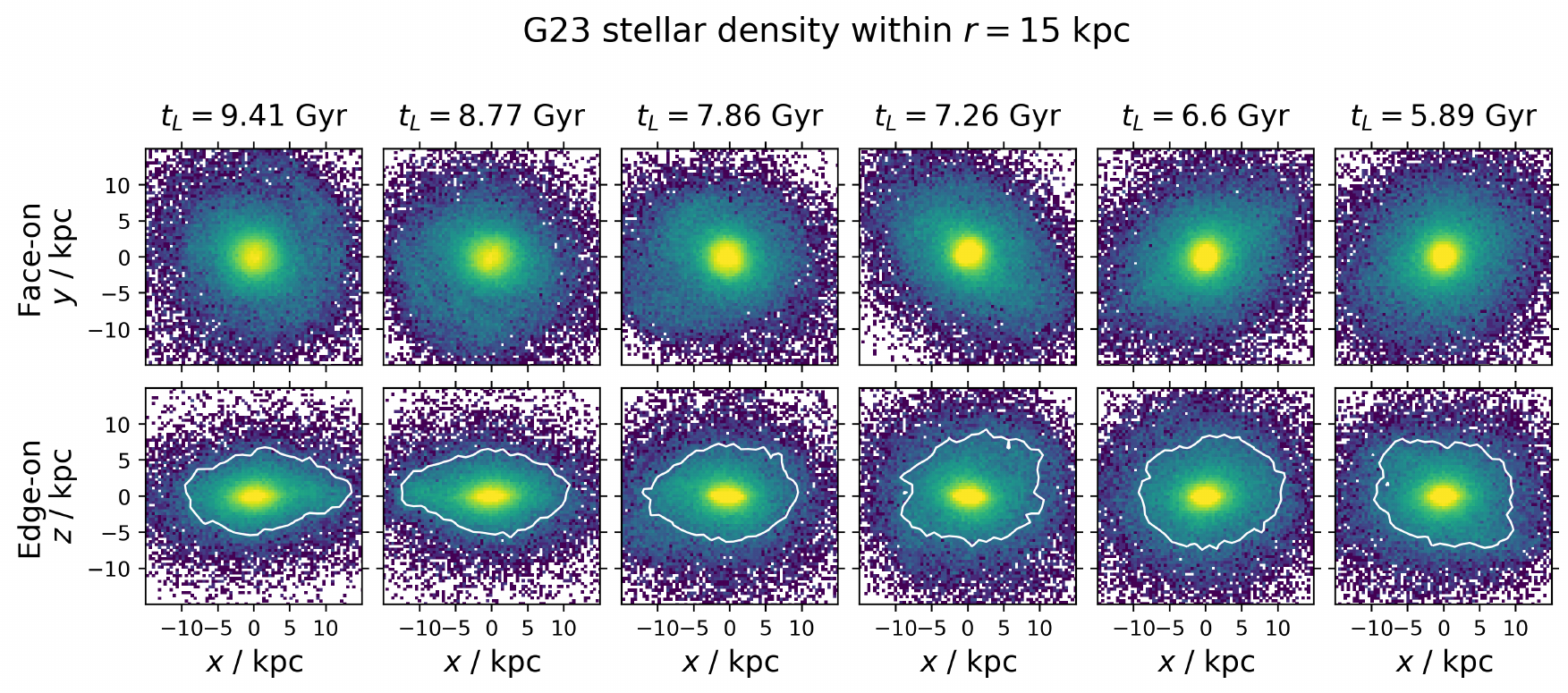}
    \caption{Density of `Splash' star particles in G23 across several snapshots. The top row shows projections along the AM axis of the disc (re-calculated in each snapshot), while the bottom row shows edge-on views. The MMAP merger occurs around the snapshot at $t_L=7.86$ Gyr. In the bottom row it can be seen that the distribution of these stars becomes less flattened along the axis of the disc after this snapshot, at least for $r>5$~kpc. This is emphasised by the white contours, which mark a fixed projected density in the $x-z$~plane. This is consistent with the idea that the merger disrupts disc stars onto more extreme orbits.}
    \label{fig:G23_splash}
\end{figure*}

This indicates that there is a relation between the orbit of the satellite before the merger and the AM flipping. On average the flipping acts to increase the alignment between the AM of the galaxy and the MMAP's pre-merger OAM. In particular, the preference for $\mathrm{cos}\:\alpha_\mathrm{after}$ and $\mathrm{cos}\:\alpha_\mathrm{today}$ to be positive suggests that the \emph{sign} of the OAM along its axis is important, not merely the orbital plane itself.\par
This does not necessarily imply that the MMAP merger itself is responsible for the realignment. Other accreted matter approaching from the same direction may share the same orbital plane, also causing realignment. However, previous work gives reason to believe that these major mergers are the most important. \citet{Welker} found in simulations that steady accretion along a filament with no mergers results in the galaxy's spin being aligned \emph{with} the filament (i.e., perpendicular to any OAM of the infalling mass). They found that it is the mergers along these filaments which cause realignment of the spin perpendicular to this direction. That is consistent with what is seen in \texttt{ARTEMIS}. If a satellite is approaching the host along a filament, its OAM will necessarily be perpendicular to that filament. Fig.~\ref{fig:orbital_AM_alignment} therefore suggests a preference for realignment of the host's AM perpendicular to this filament. Note however that in most cases the alignment remains poor after the merger. In more than half of galaxies $\mathrm{cos}\:\alpha_\mathrm{after}<0.5$, so the angle between the axes is greater than $60^\circ$. The alignment is only \emph{increased} after the merger, but far from perfect.\par
Note also that those galaxies with closer alignment \emph{before} the merger (i.e., larger $\mathrm{cos}\:\alpha_\mathrm{before}$) tend to experience a smaller change in angle of their AM vectors during and after their mergers. This can be seen in Fig.~\ref{fig:orbital_AM_alignment}; points on the right-hand sides of each panel ($\mathrm{cos}\:\alpha_\mathrm{before}>0$) mostly have small angle changes of $\lesssim50^\circ$ (with the obvious exception of G29). Those on the left undergo much larger changes of orientation, especially where the angular momenta are highly misaligned ($\mathrm{cos}\:\alpha_\mathrm{before}\lesssim-0.5$). This suggests that the \emph{amount} of flipping experienced is related to the initial configuration of the system. The host may be more likely to be significantly flipped if the satellite's orbital inclination with respect to the disc is large.\par
The right-hand panel of Fig.~\ref{fig:orbital_AM_alignment} may suggest a possible method of constraining the orbital plane of the GS progenitor before its merger with the Milky Way. Based on this evidence, it appears unlikely that the present-day disc AM axis is significantly misaligned (by more than $90^\circ$) from the former orbital angular momentum axis of the GS progenitor.\par
It is worth considering whether the OAM of a satellite could significantly affect the AM of a disc on an order-of-magnitude basis. The magnitudes of the satellite OAM and disc AM are of order, respectively,
\begin{align}
    J_\mathrm{orb, sat} &\sim M_\mathrm{sat}r_\mathrm{sat}v_\mathrm{sat}, \\
    J_\mathrm{disc} &\sim M_\mathrm{disc}R_\mathrm{disc}v_\mathrm{circ},
\end{align}
where $M_\mathrm{sat}$ and $M_\mathrm{disc}$ are the total masses of the satellite and disc, $r_\mathrm{sat}$ is the distance between the satellite and the host, $R_\mathrm{disc}$ is a disc scale-length, $v_\mathrm{sat}$ is the tangential orbital velocity of the satellite, and $v_\mathrm{circ}$ is the circular velocity in the disc. Typical values of these quantities prior to a merger are:
\begin{itemize}
    \item $M_\mathrm{sat}\sim10^{11}\, {\rm M}_\odot$, $M_\mathrm{disc}\sim10^{10}\,{\rm M}_\odot$,
    \item $r_\mathrm{sat}\sim100$ kpc, $R_\mathrm{disc}\sim 5$~kpc,
    \item $v_\mathrm{sat}\sim100$ kms$^{-1}$, $v_\mathrm{circ}\sim 200$ kms$^{-1}$,
\end{itemize}
using values from the merger trees and \citet{Font}. The ratio of angular momenta is then
\begin{equation}
    \frac{J_\mathrm{orb, sat}}{J_\mathrm{disc}}\sim\frac{M_\mathrm{sat}}{ M_\mathrm{disc}}\frac{r_\mathrm{sat}}{R_\mathrm{disc}}\frac{v_\mathrm{sat}}{v_\mathrm{circ}}\sim100.
\end{equation}
The satellite's OAM is therefore roughly two orders of magnitude larger than the disc AM, and is more than capable (in principle) of causing a spin flip. In reality, much of the OAM will likely be transferred to the host's DM halo instead of the disc. However, this calculation demonstrates that it is not unreasonable that the OAM might play an important role in the process of AM flipping.

\begin{figure}
    \centering
    \includegraphics[width=\columnwidth]{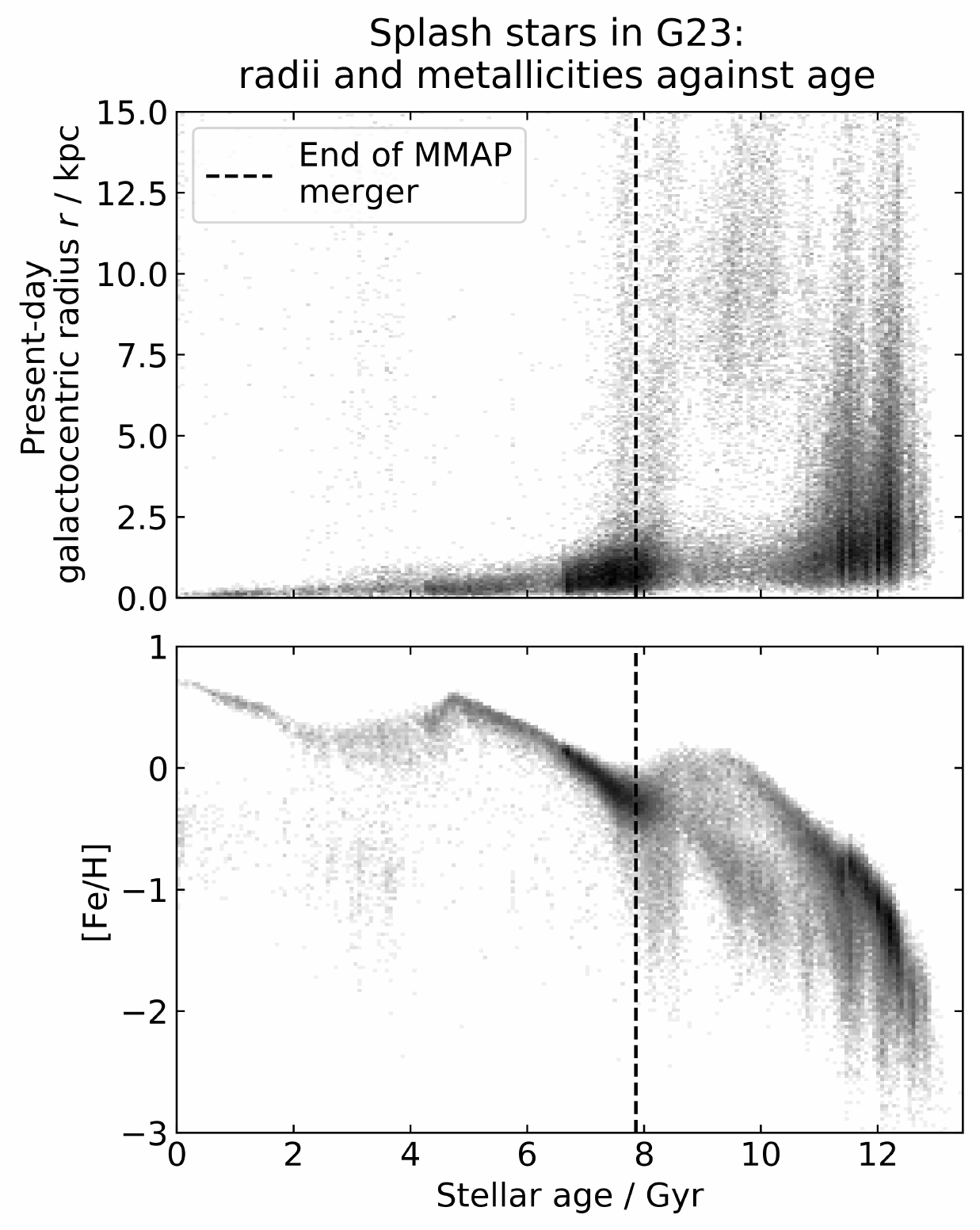}
    \caption{\textbf{Top panel:} Present-day radii and ages of Splash stars in G23, with the approximate MMAP merger time marked. There are two major episodes of formation of these stars; the later one at 8 Gyr ago coincides with the merger and produces stars mostly with radii $r<2$~kpc at the present-day. Stars born in the earlier star formation period are much less centrally concentrated. \textbf{Bottom panel:} Metallicity [Fe/H] of the same sample of stars, again plotted against age. As expected, younger stars are generally more metal-rich. However, the burst of star formation at 8 Gyr is accompanied by a decrease in the average metallicity of new stars.}
    \label{fig:G23_splash_ages_radii}
\end{figure}

\subsection{Disc splashing}
\label{section:splashing}
The `Splash' is a population of metal-rich stars in the Milky Way on extreme orbits, with little angular momentum \citep{splash}. These stars are believed to be on their present-day orbits due to the GS merger. In this section we discuss analogues of the Splash in the Sample~GS/discs \texttt{ARTEMIS} galaxies, and investigate their origins.\par
 \citet{splash} found that the cleanest selection cut for Splash stars in the Milky Way is $v_\theta<0$ and [Fe/H] $>-0.7$, where the lower limit on [Fe/H] is due to the GS being dominant at lower metallicities. Inspired by this, in each galaxy we selected samples of \textit{in situ} star particles with $v_\theta<0$. These samples contain stars which were born bound to the host subhalo (at any radius), \textit{and} are on retrograde orbits at the present-day. Any particles born bound to lower mass subhaloes (i.e. accreted star particles) are therefore excluded. We dispensed with the [Fe/H] cut due to the availability of \textit{in situ} / accreted information. This also removes any bias caused by differences in metallicity distributions between different galaxies. The purpose of these samples is to provide analogues of the Splash sample in the Milky Way. Henceforward these will be referred to as `Splash' samples. Note however that they will likely contain significant numbers of retrograde stars belonging to a bulge at small radii.\par
 
 As an example, the stellar density of the Splash sample in G23 is plotted across several snapshots in Fig.~\ref{fig:G23_splash}. The two projections are face-on and edge-on to the disc, with the disc orientation re-calculated in each snapshot to take disc flipping into account. The most obvious changes occur in the period between lookback times of 8.77~Gyr and 7.26~Gyr, which corresponds to the time of the MMAP merger. At radii beyond $r\sim5$~kpc the distribution becomes more spherical, less flattened along the $z$-axis. In addition, the stellar density at small radii increases significantly over the same time period. This rapid burst of star formation can be seen more clearly in the top panel of Fig.~\ref{fig:G23_splash_ages_radii}, which plots the present-day radii of the retrograde \textit{in situ} stars against their age. These stars are mostly born in two distinct periods. Those formed before $\sim10$~Gyr are spread over a wide range of radii today, extending to around 10~kpc. By contrast, the younger population of stars with ages between 6.5 and 8.5~Gyr largely occupies the inner regions of the galaxy at the present-day, within a radius of $\sim2$~kpc. The birth of this younger population coincides with the MMAP merger. The lower panel shows that these stars typically have a lower metallicity than those born a Gyr earlier. This is therefore likely to be a merger-induced starburst, with stars born partially from the MMAP's metal-poor gas.

\subsubsection{Origin of the Splash populations}
To determine the origin of the retrograde stars we traced them back to earlier snapshots. We divided each sample into two, by whether present-day radii are greater than or less than 5~kpc. The sub-sample with radii $r>5$~kpc is of more interest, since it roughly corresponds to the Splash population in the solar neighbourhood studied by \citet{splash}. It also excludes any stars belonging to a bulge, and starburst populations at small radii as in G23. We found the age distributions of both sub-samples, and measured the direction of the net AM vector of the $r>5$~kpc sub-sample as a function of time. \par

\begin{figure*}
    \centering
    \includegraphics[width=0.8\textwidth]{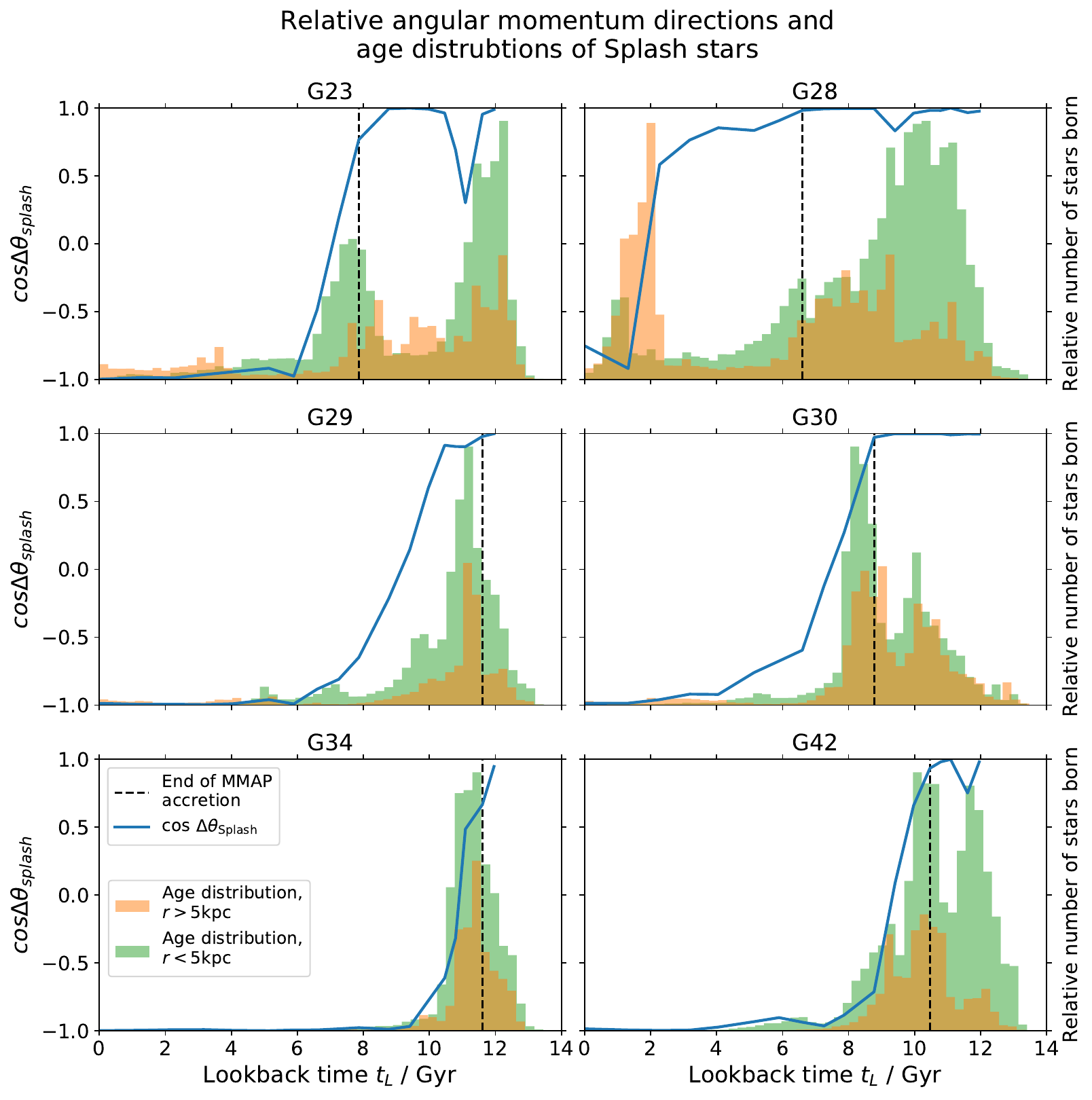}
    \caption{Alignment between AM axes of \textit{in situ} and Splash star particles at $r>5$~kpc at the present day (blue traces), for six different Sample~GS/discs galaxies. Age distributions are plotted for these stars (orange histograms) and also for the Splash stars at $r<5$~kpc at the present-day (green histograms). For each galaxy the two histograms share the same normalisation. The black dashed lines mark the approximate times at which the MMAPs were finally accreted. Initially the axes are well-aligned, with $\mathrm{cos}\:\Delta\theta_{\mathrm{splash}}\approx1$. Reversals in the relative axis directions are then seen in all six cases, after which few new stars are born. Apart from in G28, these reversals approximately coincide with the MMAP mergers. The age distributions take different forms in each case. There is often a peak in the distributions coincident with the MMAP mergers, which may be due to a merger-induced starburst.}
    \label{fig:angle_splashes}
\end{figure*}

\citet{splash} argued that Splash stars in the Milky Way were largely members of a proto-disc prior to the GS merger. If this is the case in the \texttt{ARTEMIS} galaxies, the net AM direction of the Splash analogue samples should be closely aligned with that of the \textit{in situ} stars as a whole. Let $\Delta\theta_{\mathrm{splash}}$ be the angle between AM vector directions of a) all \textit{in situ} stars within $r=30$~kpc, and b) all stars in the Splash sub-sample with $r>5$~kpc. Then $\mathrm{cos}\:\Delta\theta_{\mathrm{splash}}=1$ for parallel spins, and $-1$ for anti-parallel spins. Note that by construction, $\mathrm{cos}\:\Delta\theta_{\mathrm{splash}}$ will be negative at the present-day, since the Splash stars are all retrograde. \citet{splash} also predict that the end of Splash star formation should be roughly coincident with the time of the merger. This information is shown in Fig.~\ref{fig:angle_splashes}, which plots $\mathrm{cos}\:\Delta\theta_{\mathrm{splash}}$ as a function of time, along with the Splash age distributions. The age distributions of the $r<5$~kpc sub-samples are also plotted for comparison. Six different galaxies are shown, which have a range of MMAP mass ratios and merger times. The lower four panels show the galaxies with early, high mass ratio mergers.\par

In all six cases the spin axes are closely aligned at some point in the past, with \hbox{$\mathrm{cos}\:\Delta\theta_{\mathrm{splash}}\approx1$.} The bulk of Splash star formation takes place while this is the case. The relative AM directions are then reversed in time intervals of $\sim2$ Gyr, and remain almost anti-parallel until the present-day. The Splash star formation rate decreases markedly at the same time, with only a small fraction of Splash stars born after the reversal. In some cases the star formation continues for $\sim1$ Gyr after the end of accretion. In Fig.~\ref{fig:angle_splashes} this can be seen clearly in G30 and G42. In all six galaxies shown there are significant peaks in star formation coincident with the reversals of relative AM.\par

In some cases (e.g., G30), the star formation histories for stars inside and outside $r=5$~kpc are similar. In others there are significant differences, which vary between different galaxies. As already discussed, in G23 there is a peak in star formation coincident with the MMAP merger and relative AM reversal which produces stars mainly at small radii. However, the reverse is true in G28; while there is again a peak in star formation at the time of the reversal, in this cases it is dominated by stars at $r>5$~kpc. There is therefore great diversity in the star formation histories of retrograde populations.\par

\begin{figure*}
    \centering
    \includegraphics[width=0.8\textwidth]{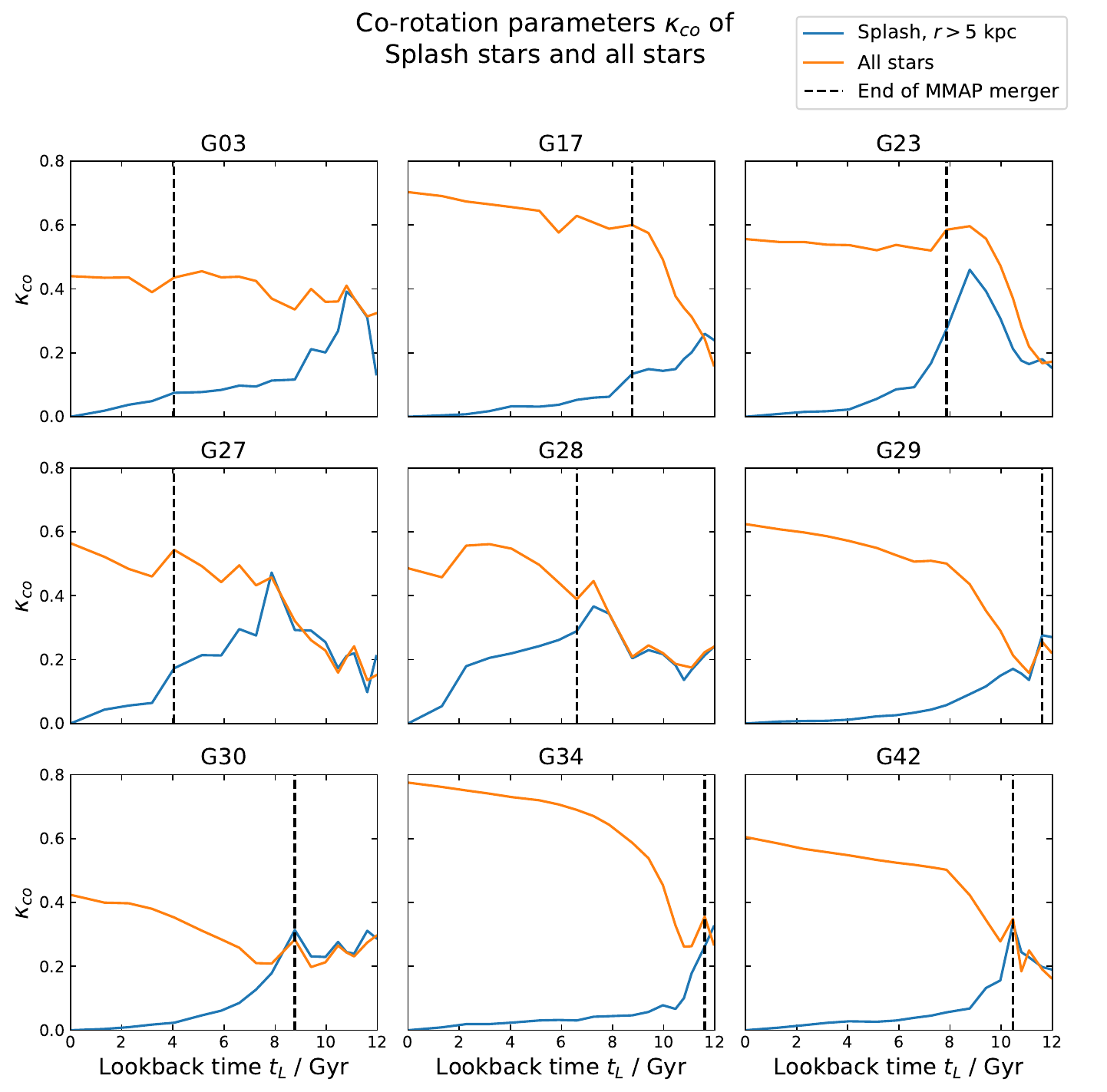}
    \caption{Co-rotation parameters $\kappa_\mathrm{co}$ for Splash stars (blue) and all stars (orange) within $r=30$ kpc, for nine Sample~GS/discs galaxies. The value of $\kappa_\mathrm{co}$ is generally lower for the Splash stars because these are all retrograde at the present day. In G28 and the four high mass ratio cases (G29, G30, G34 and G42), the values of $\kappa_\mathrm{co}$ for each population are approximately equal prior to the MMAP merger. In the other galaxies the values diverge at an earlier time. In some cases $\kappa_\mathrm{co}$ decreases for both populations at around the time of the merger.}
    \label{fig:kappa_splashes}
\end{figure*}

In G23, G29, G30, G34 and G42 the reversals in relative AM direction and star formation peaks are within $~1 Gyr$ of the MMAP mergers, while in G28 the merger pre-dates the reversal by 4 Gyr. However, checking the merger tree for G28 reveals that another merger took place with a satellite of peak stellar mass $\sim5\times10^{10}\, {\rm M}_\odot$ at $t_L\sim2$~Gyr, coincident with the reversal. Hence in each of these four cases there is a merger with a massive satellite taking place at the same time as the reversal of relative AM directions (and an associated starburst).\par
Of the fifteen Sample~GS/discs galaxies:
\begin{itemize}
    \item{11 show similar reversals within $\sim1$ Gyr of the MMAP merger event,}
    \item{3 (G03, G28 and G43) show reversals at other times, all of which coincide with the accretion of other massive satellites (stellar masses $\gtrsim5\times10^8\,{\rm M}_\odot$),}
    \item{G44 does not show such a reversal.}
\end{itemize}
The four high mass ratio cases do not here show any notable difference in behaviour compared with the other galaxies.
These plots indicate that the MMAP (or other) mergers are important in creating these retrograde populations. Only a small proportion of these stars are born after the reversal. This suggests that there is no process happening at later times which is able to place significant numbers of stars onto retrograde orbits.\par
However, this is not sufficient to show that the retrograde populations are due to a \emph{single} merger event. For example, stars born earlier could be scattered onto retrograde orbits by earlier mergers. This would not be evident in the net AM direction if the angular momentum was dominated by prograde stars at a given time. To test whether the Splash stars were part of a disc (or proto-disc) prior to the MMAP merger, we calculated $\kappa_\mathrm{co}$ as a function of time for the Splash samples (see eq.~\ref{equation:kappa}). The net stellar AM axis of each galaxy was recalculated in each snapshot and used as the reference axis. Again, we included only Splash star particles with $r>5$ kpc at the present-day to exclude bulge stars. Note however that this does not necessarily exclude Splash stars with $r<5$ kpc at earlier times. Examples are plotted in Fig.~\ref{fig:kappa_splashes} including the same galaxies as above, with the corresponding traces for all stars for comparison. Note that due to the selection cuts, the values at the present-day are forced to be $\kappa_\mathrm{co}=0$ (Splash) and $\kappa_\mathrm{co}>0.4$ (all).\par

\begin{figure*}
    \centering
    \includegraphics[width=0.8\textwidth]{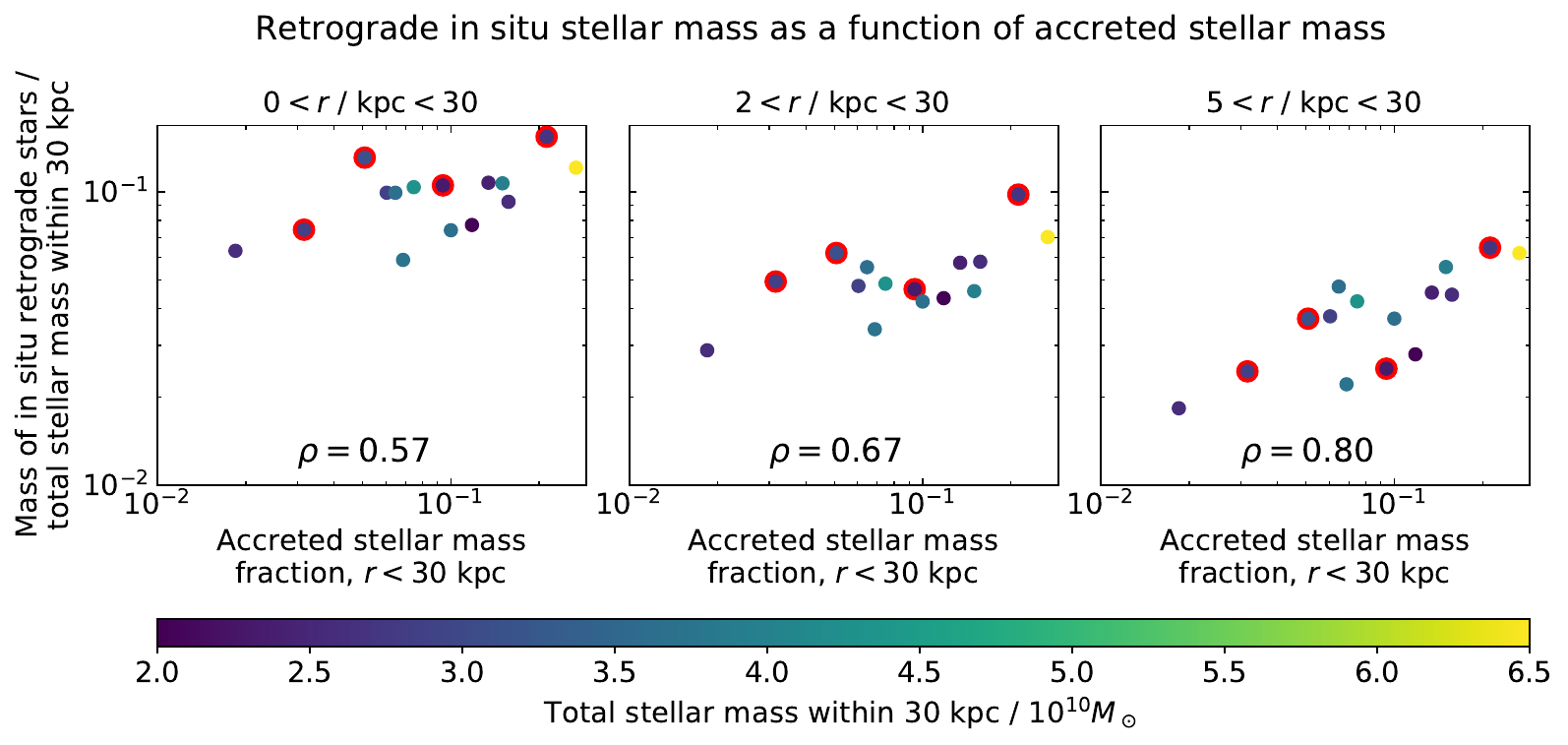}
    \caption{Mass of retrograde \textit{in situ} stars against accreted stellar mass, in all Sample~GS/discs galaxies. The points highlighted in red correspond to G29, G30, G34 and G42. The accreted stellar mass is measured within $r=30$ kpc, and both quantities are plotted as fractions of the total stellar mass within $r=30$ kpc (colour-coded). The retrograde mass is measured in the radial intervals $0<r/$kpc $<30$ (left panel), $2<r/$kpc $<30$ (middle panel) and $5<r/$kpc $<30$ (right panel). In the former two cases a significant proportion of the star particles are likely to belong to the bulge rather than an analogue of the Milky Way's Splash. The vertical shift in the points between the panels is due to the extra mass of retrograde stars at small radii. In all three plots a weak correlation between accreted stellar mass and retrograde \textit{in situ} mass is discernible. This is confirmed by the Pearson correlation coefficients $\rho$, which are shown in each panel. The correlation becomes weaker as the inner radial cut is relaxed.}
    \label{fig:splash_correlation}
\end{figure*}

In all nine galaxies shown in Fig.~\ref{fig:kappa_splashes} the values of $\kappa_\mathrm{co}$ for the Splash and overall populations are approximately equal at early times, before they diverge. This suggests that the Splash stars are initially representative members of the stellar population as a whole, in that their rotational properties are very similar (Fig.~\ref{fig:angle_splashes} also demonstrates that the AM axis  is aligned with that of \textit{in situ} stars). In G28, G29, G30, G34 and G42 the times at which the traces diverge correspond closely to the times of the MMAP mergers. It is notable that the latter four of these are the galaxies which experienced the early high mass ratio mergers. These plots suggest that the mergers do perturb the kinematics of the \textit{in situ} stars of these galaxies, and contribute to their Splash populations. Note that G29 and G42 both exhibit extremely rapid swings of angular momentum in Fig.~\ref{fig:angle_rates}. Such large changes in the overall kinematics of the \textit{in situ} populations are likely linked to the perturbation of some of these stars onto retrograde orbits. The traces in G23 also move apart significantly when the merger happens. In each of these cases, the MMAP mergers appear to be the first events which significantly altered Splash kinematics compared to the other stars. This is unsurprising, given that most of these mergers happened relatively early. Interestingly, G28 is a case where the total AM direction does not undergo a reversal at the time of the MMAP merger, as discussed in relation to Fig.~\ref{fig:angle_splashes}. Despite this, the MMAP clearly has a significant effect on the rotation of the Splash population.\par
In the other galaxies the values of $\kappa_\mathrm{co}$ for each population are already different by the time of the MMAP merger. This is the case in most of the fifteen Sample~GS/discs galaxies. In G27 the values diverge sharply at $t_L=8$ Gyr ago, which corresponds to the merger time of a satellite of peak total mass $5\times10^{10}\,{\rm M}_\odot$. This suggests that at least some of the present-day retrograde stars were `splashed' onto more extreme orbits by this earlier merger, not just that of the MMAP. Similar conclusions also apply to the other galaxies. The age distributions show that a significant proportion of Splash stars have already been born before the MMAP merger, so the different values of $\kappa_\mathrm{co}$ are not an artefact of measuring a small number of stars. We have also checked that the fractions of stars in the Splash populations are small, so the similar values of $\kappa_\mathrm{co}$ at early times are not the results of measuring the same population twice.\par
In the majority of the galaxies shown, the values of $\kappa_\mathrm{co}$ decrease at the MMAP merger times, for both populations. This is also seen in G28 at the later merger time of $t_L=2$ Gyr. For the full stellar populations the decreases last for less than $\sim1$ Gyr, before $\kappa_\mathrm{co}$ either increases again (e.g., in G27) or remains stable (e.g., G23). This is likely due to disc growth: newly born stars in a disc will tend to have high $\kappa_\mathrm{co}$ (by definition of a disc). For the Splash population, $\kappa_\mathrm{co}$ continues to slowly decrease to zero at the present-day. This might be explained by the changing orientation of the disc. If the orientations of the Splash and disc are not tightly coupled, $\kappa_\mathrm{co}$ could decrease as a result of the disc slowly precessing relative to the Splash.\par
In summary, Figs.~\ref{fig:angle_splashes} and \ref{fig:kappa_splashes} lend support to the conclusion of \citet{splash} that the present-day retrograde stars are on their extreme orbits due to mergers. In almost all cases their net AM direction with respect to the galaxy is reversed at times corresponding to significant mergers. The co-rotation parameters $\kappa_\mathrm{co}$ of the Splash populations also decrease compared to the rest of the stars. The Splash populations therefore become less disc-like and more retrograde. However, it is clear that in many of these galaxies the MMAP is not solely responsible for the Splash populations, with earlier events frequently causing disruption.

\subsubsection{Connection to accreted stellar mass}
\citet{splash} used the youngest age of Splash stars in the Milky Way to calculate the epoch of the GS merger. This is supported by Fig.~\ref{fig:angle_splashes}, which shows sharp decreases in the Splash star formation rates at the same times as the mergers (see discussion above). The authors also proposed using the Splash to calculate properties of the GS progenitor, such as mass and merger geometry. This has already been explored in the Auriga simulations by \citet{mash}, who showed that the fraction of \textit{in situ} stars on retrograde orbits (within certain metallicity cuts) positively correlates with the stellar mass ratio of the GS merger. Since the Splash analogues in our selection of \texttt{ARTEMIS} galaxies are likely created by multiple mergers, a better correlation may exist between the Splash fraction and the \emph{total} accreted stellar mass. Fig.~\ref{fig:splash_correlation} plots the mass of retrograde \textit{in situ} stars against the total mass of accreted stars within 30 kpc, for all fifteen Sample~GS/discs galaxies. All values are measured at the present-day and plotted as fractions of the total stellar mass within 30~kpc. The retrograde mass is measured within three different apertures: $0<r/$kpc~$<30$, $2<r/$kpc~$<30$ and $5<r/$kpc~$<30$. The purpose of the radial cuts is to focus on the `solar neighbourhood' and exclude stars at small radii whose presence may result from different processes (e.g., starbursts). All other masses are measured in the aperture $0<r/$kpc~$<30$. The points are colour-coded by total stellar mass of the host within 30 kpc (at the present-day). This is to demonstrate that the distribution of points is not merely a result of scaling by this quantity: there is no apparent correlation between this and the positions of the points. Since correlations in these plots could also result from different galaxies having different scale-lengths, we also checked that there was no correlation between the galaxies' half-mass radii and any of the plotted quantities.\par
In all three plots there is a visible correlation between accreted stellar mass fraction and retrograde \textit{in situ} mass fraction, confirmed by the reasonably high Pearson correlation coefficients $\rho$. This is expected, since the greater the accreted mass, the greater the mass of stars which can scattered onto retrograde orbits. This confirms the findings of \citet{mash}. However, the correlation becomes weaker ($\rho$ decreases) in the middle and left-hand panels as the inner limit on $r$ is reduced. As suggested by Figs.~\ref{fig:G23_splash_ages_radii} and \ref{fig:angle_splashes}, this may be partly due to the presence of stars at small radii formed in merger-induced starbursts (as seen in G23) which are not \textit{scattered} onto their orbits by merging satellites. The considerable scatter in the data points could also be due to many factors not considered here; for example \citet{mash} reported correlations with the orbital eccentricity of the merger. However, these correlations do suggest that measuring the mass of the Milky Way's Splash could be a possible route to constraining the mass of the GS progenitor.

\section{Conclusions}
\label{sec:concl}
This work uses the \texttt{ARTEMIS} set of high-resolution zoomed-in galaxy formation simulations to study the merger histories of Milky-Way like galaxies. Motivated by the discovery of the \textit{Gaia} Sausage (GS) merger in the Milky Way, we have focused on the transformations experienced by the host galaxies caused by similar accretion events. This follows on from previous studies \citep[e.g.,][]{Fattahi, Mackereth2019,Bignone2019,splash, Elias2020,mash,Renaud2021} which have used other cosmological simulations to explore the effects of such mergers, including the formation of a Splash. However, we have also addressed some phenomena (e.g., disc flipping) which have not been extensively studied specifically in the context of the GS-Milky Way interaction. Our principal findings are summarised below.\par
\textbf{(i) Prevalence of GS-like features in Milky Way-mass galaxies.} By plotting the velocity distributions of accreted star particles in spherical polar coordinates, we found radially anisotropic features resembling the GS. Using a Gaussian mixture model to fit these distributions, we characterised the GS-like components by their anisotropy $\beta$ and their contribution, following \citet{Fattahi}. This shows that GS analogues are very common; $\approx1/3$ or more of the \texttt{ARTEMIS} galaxies contain a similarly anisotropic and dominant feature, depending on the selection cut used. This fraction is unchanged when only disc galaxies are considered. In a selection of fifteen disc galaxies containing GS analogues (Sample~GS/discs), we traced the star particles from the accreted satellite with highest stellar mass (MMAP). The total MMAP stellar masses are on the order of $10^9\,{\rm M}_\odot$ in nearly all cases. Plotting the velocity distributions shows that the GS-like components are either fully or partially made up of these stars, in a large majority of the Sample~GS/discs galaxies. This is consistent with the conclusion of a number of previous studies \citep[e.g.][]{Deason_break,Belokurov,Helmi} that the GS in the Milky Way is the debris from a merger with a single massive satellite.\par
\textbf{ii) Mass ratios and times of most massive merger events.} The total mass ratios of the MMAP mergers span an order of magnitude, between $\sim0.05$ and $\sim0.6$. For a total host mass of order $10^{12}\,{\rm M}_\odot$, the lower value corresponds to a satellite with total mass greater than $10^{10}\,{\rm M}_\odot$. This is consistent with the mass estimates available in the literature \citep[][]{Belokurov,Helmi,Fattahi,Myeong}. The merger times also span a wide range, between lookback times of 2 and 12 Gyr. However, there is a preference for these mergers to be earlier, with only 3 of the 15 occurring in the 5 Gyr before the present-day. 10 of the 15 have accumulated more than $90\%$ of their accreted stellar mass within $r=30$~kpc before a lookback time of 5 Gyr, while only two have gained less than $60\%$. Note that GS-like major merger events, e.g. those with mass ratios above 0.4, corresponding to the estimate found by \citet{Naidu_reconstruction}, are limited to lookback times greater than 8 Gyr \citep[in line with the age dating of the merger by e.g.][]{Helmi,Dimatteo2019,Gallart2019,Vincenzo2019,splash,Bonaca2020,Das2020,Chaplin2020,Montablan2021}. 
\par
\textbf{iii) Disc angular momentum evolution.} We found the stellar angular momentum (AM) axis directions of the host galaxies in each simulation snapshot. These directions are highly variable as a function of time, and many of the galaxies have experienced large changes in orientation since the time of the MMAP merger. This suggests that attempts to reconstruct details of the GS merger in the Milky Way \citep[e.g.][]{Villalobos2008,Koppelman2020,Naidu_reconstruction,Vasiliev_radial} should take these possible changes into account.\par
There are two apparently distinct regimes of AM axis variation observed: a) rapid flips of up \hbox{to $\sim90^\circ$} on timescales comparable to an orbital period, coincident with massive mergers; and b) slower precession at rates on the order of \hbox{10 degrees Gyr$^{-1}$}, even when there is no significant ongoing accretion of stars. There are multiple cases of rapid flips (e.g., G19 and G23) where the co-rotation parameter $\kappa_\mathrm{co}$ remains large enough for the host galaxy to be considered a disc throughout the flip. These conclusions are unchanged when only stars born before the flip are considered. This is strong evidence that stellar discs can survive rapid AM flips. Defining a disc flip to be a change of $60^\circ$ or more in a period of 2~Gyr (throughout which $\kappa_\mathrm{co}>0.4$), we found that four of the fifteen Sample~GS/discs galaxies experience such a flip, all of which coincide with the MMAP mergers in these galaxies.\par
We calculated the maximum rates of AM flipping in intervals of a few Gyr around the times of the MMAP mergers. Rates of up to \hbox{$\approx60$ degrees Gyr$^{-1}$} are measured for the galaxies whose stars form a disc at that time. These rates appear to be uncorrelated with the total mass ratio of the merger, and significant flips occur even when the mass ratio is only $\sim0.05$ (e.g., in G19).
It remains to be determined whether the Milky Way itself experienced such a flip. For this purpose some signature of flipping observable in the present-day galaxy is required, such as in the phase space of old stars. Therefore, a natural continuation of this work would be to search for such a signature in simulated galaxies.\par
\textbf{iv) DM halo metamorphosis.}
The minor axes of the dark matter (DM) haloes measured out to different radii are similarly variable through time. In some cases (e.g., G23) they also undergo rapid changes of orientation around the times when MMAP mergers occur. Often, e.g. in G23, this change occurs before the corresponding disc flip, suggesting that the DM halo torques the inner galaxy into its new configuration. We also find that DM haloes can be significantly re-shaped by events involving the accretion of GS-like satellites, with 7/15 of the Sample~GS/discs galaxies experiencing a change in sphericity of more than 0.2 Gyr within a 2~Gyr period around the MMAP merger time. It is likely that this DM halo metamorphosis is not caused by the GS merger itself but is contemporaneous with it. The GS progenitor is simply the largest block of matter accreted onto the MW at the time, with plenty more material, both smooth (DM and gas) and lumpy (smaller dwarf galaxies and DM sub-halos), arriving from the same direction. We show that most of this matter settles in the outer Galaxy, beyond 30 kpc from its centre.\par
\textbf{v) Host orientation at $z=0$.}
Using merger trees, we determined the orbital angular momentum (OAM) of the MMAP about the host prior to the merger. We found the alignment of the host's AM axis with this direction a) before the merger, b) soon after the merger and c) at the present-day. This shows that on average the host AM axis becomes more aligned with the OAM after the merger. By the present-day the alignment has increased further in most cases; the present-day disc AM vector points within $90^\circ$ of the MMAP's former OAM in all but one case. This implies a relation between the AM flips and the merger geometry. Since the magnitude of the satellite's OAM is typically two orders of magnitude larger than the disc's AM, this connection is unsurprising.\par
\textbf{vi) Disc splashing.} We investigated the histories of samples of \textit{in situ} stars which are on retrograde orbits at the present-day (`Splash' samples). In the vast majority of galaxies the net AM direction of these stars is aligned with that of the full \textit{in situ} population at early times. Then in the space of a few Gyr these AM vectors become highly misaligned, with the times of these reversals coincident with accretion of massive satellites. These mergers are often those with the MMAPs. This supports the conclusion of \citet{splash} that the Splash population in the Milky Way consists of \textit{in situ} stars whose orbits were highly altered by a massive merger event \citep[see also][]{Bonaca2017,Gallart2019,Dimatteo2019,Bonaca2020}. In most cases only a small proportion of these stars are born after this reversal. This adds to the evidence of \citet{Dimatteo2019}, \citet{mash} and \cite{splash} that the age distributions of counter-rotating \textit{in situ} populations can be used to constrain the merger times.\par
We measured the co-rotation parameter $\kappa_\mathrm{co}$ (see eq.~\ref{equation:kappa}) of each Splash sample, and show that at early times it matches the value for the full stellar population in a large majority of cases. This is again consistent with the notion that Splash stars were originally on more typical orbits. However, in most cases $\kappa_\mathrm{co}$ of the Splash sample decreases below the overall value \emph{earlier} than both the MMAP merger and the relative AM reversal. This implies that the Splash population becomes less `discy' than the host galaxy as a whole before this merger. Hence the MMAP mergers are not the only events responsible for altering the orbits of the present-day retrograde stars. This suggests that the Splash in the Milky Way could have been created by mergers with multiple satellites, not just the GS progenitor. This should be taken into account when attempting to infer information about the GS merger from properties of the Splash.\par
We also show that there is a loose correlation between total mass (ratio) of \textit{in situ} retrograde stars and total mass (ratio) of accreted stars. This is similar to the relation found by \citet{mash} in the Auriga simulations, between retrograde mass fraction and stellar mass of a massive accreted satellite. This correlation is best revealed when retrograde stars in the central regions are excluded. We see evidence that these are partially stellar populations formed in star-bursts triggered by the MMAP accretion \citep[in agreement with][]{mash}. We demonstrate that poorly-enriched gas supplied during the mergers is used in these induced star-formation events. \par
We have demonstrated that the transformation of our Galaxy due to and during the ancient massive merger is not limited to the previously known phenomena such as heating, scrambling and splashing of the MW disc.  We see that GS-like events are coincident with disc flipping and triggering of star-formation. To quantify the exact conditions for each of these and to reconstruct the exact sequence of the changes that produced the MW as we know it, a larger and more sophisticated set of numerical simulations of galaxy formation is needed.  We look forward to comparing the future Galaxy models with the chemical and chronological constraints from the new wide-area spectroscopic surveys such as DES, WEAVE and 4MOST as well as the future data releases of {\it Gaia}.

\section*{Acknowledgements}
The manuscript has been improved thanks to the helpful suggestions of an anonymous referee. The authors thank John Helly for generating the merger trees for the \texttt{ARTEMIS} simulations. They are also grateful to Romeel Dav\'e, Katarina Kraljic, Rosemary Wyse, Eugene Vasiliev for many illuminating discussions that have inspired some of the investigations discussed here. VB diu moltes gr\`acies a la Universitat de Barcelona per acollir-lo.

This research made use of Astropy,\footnote{http://www.astropy.org} a community-developed core Python package for Astronomy \citep{astropy:2013, astropy:2018}.

This project has received funding from the European Research Council (ERC) under the European Union's Horizon 2020 research and innovation programme (grant agreement No 769130).  This work used the DiRAC@Durham facility managed by the Institute for Computational Cosmology on behalf of the STFC DiRAC HPC Facility. The equipment was funded by BEIS capital funding via STFC capital grants ST/P002293/1, ST/R002371/1 and ST/S002502/1, Durham University and STFC operations grant ST/R000832/1. DiRAC is part of the National e-Infrastructure.

\section*{Data Availability}

The data underlying this article may be shared on reasonable request to the corresponding author.



\bibliographystyle{mnras}
\bibliography{refs} 



\appendix
\FloatBarrier
\section{Table of parameters}\label{sec:table}
Table~\ref{table:params} presents some of the most important quantites used in this study, for the fifteen haloes in Sample~GS/discs.
\begin{table}
    \centering
    \begin{tabular}{lrrrr}
    \hline
     Halo   &   \makecell{Present-day\\$\kappa_\mathrm{co}$} &   \makecell{MMAP merger\\total\\mass ratio} &      \makecell{MMAP peak\\stellar\\mass / $10^9M_\odot$} &      \makecell{MMAP merger\\lookback\\time / Gyr} \\
    \hline
     G01    &                               0.61 &                    0.2  & 1.97 &  8   \\
     G03    &                               0.44 &                    0.11 & 4.25 &  4.1 \\
     G17    &                               0.7  &                    0.16 & 1.9  &  8.8 \\
     G19    &                               0.68 &                    0.07 & 0.23 &  9.9 \\
     G23    &                               0.56 &                    0.25 & 1.48 &  7.9 \\
     G24    &                               0.53 &                    0.06 & 0.72 &  6.9 \\
     G27    &                               0.56 &                    0.26 & 3.63 &  4.6 \\
     G28    &                               0.49 &                    0.28 & 1.74 &  6.9 \\
     G29    &                               0.62 &                    0.46 & 0.98 & 11.6 \\
     G30    &                               0.42 &                    0.56 & 3.75 &  8.8 \\
     G34    &                               0.77 &                    0.43 & 0.66 & 11.6 \\
     G40    &                               0.65 &                    0.36 & 1.29 &  5.9 \\
     G42    &                               0.6  &                    0.6  & 2.03 & 10.5 \\
     G43    &                               0.45 &                    0.31 & 8.72 &  2.7 \\
     G44    &                               0.63 &                    0.1  & 1.87 &  8.8 \\
    \hline
    \end{tabular}
    \caption{The most used parameters for the Sample~GS/discs galaxies. There are a few small differences in some of the values of $\kappa_\mathrm{co}$ compared to \citet{Font}, due to a slightly different definition of $\kappa_\mathrm{co}$ used in this study.}
    \label{table:params}
\end{table}


\bsp	
\label{lastpage}
\end{document}